\documentclass[amsmath,amssymb]{revtex4}
 \usepackage{url}
\usepackage{enumerate} 
 \usepackage{graphicx}
 
\def\eqnn#1{Eq.~(\ref{eq:#1})}
\def\sect#1{Sec.~\ref{sec:#1}} 
\def\eqnb#1{(\ref{eq:#1})}
\def\eqnn#1{Eq.~\eqnb{#1}} 
\def\tabno#1{Table~\ref{tab:#1}}

\def\vev#1{\left\langle#1\right\rangle}

\def\idt#1{$\vev{p_{#1}^2}$ }
\def\nStep{N_{\rm step}}
\def\nSample{N_{\rm sample}}
\def\tPair{\tau_{\rm pair}}
\def\figno#1{Fig.~\ref{fig:#1}}
\begin{document} %
\title{Symmetry, chaos and temperature in the one-dimensional lattice
  $\phi^4$ theory} \author{Kenichiro
  Aoki\footnote{E--mail:~{\tt ken@phys-h.keio.ac.jp}.}}
\affiliation{${}^*$Research and Education Center for Natural Sciences
  and Hiyoshi Dept. of Physics, Keio University, Yokohama 223--8521,
  Japan }
\begin{abstract} 
  The symmetries of the minimal $\phi^4$ theory on the lattice are
  systematically analyzed. We find that symmetry can restrict
  trajectories to subspaces, while their motions are still chaotic.
  The chaotic dynamics of autonomous Hamiltonian systems are
  discussed, in relation to the thermodynamic laws. Possibilities of
  configurations with non-equal ideal gas temperatures in the steady
  state, in Hamiltonian systems, are investigated, and examples of
  small systems in which the ideal gas temperatures are different
  within the system are found.  The pairing of local (finite-time)
  Lyapunov exponents are analyzed, and their dependence on various
  factors, such as the energy of the system, the characteristics of
  the initial conditions are studied, and discussed. We find that for
  the $\phi^4$ theory, higher energies lead to faster pairing
  times. We also find that symmetries can impede the pairing of local
  Lyapunov exponents, and the convergence of  Lyapunov exponents.
\end{abstract}
\maketitle
\section{Introduction}
\label{sec:intro} 
Chaotic properties of Hamiltonian systems have been studied for some
time, and a general picture of the dynamics seems to be
emerging\cite{chaos1,HH1}.  Yet interesting questions from
the physics point of view still remain. 
Some of these issues were raised in \cite{HooverSnook}, which we 
address in this work.
In a dynamical system with chaos, trajectories with different initial conditions
diverge exponentially from each other in the phase space, so that it
may seem difficult to control its region of motion.  The domain within
the phase space in which it travels in can be restricted for energetic
reasons. The possibilities of the symmetries of the dynamical system limiting the
allowed region of motion is investigated in this work. The symmetries
of the minimal $\phi^4$ theory, which has no additional symmetries, is
investigated systematically. It is found that symmetries can indeed
restrict the trajectories of the dynamical system in the phase space to its lower
dimensional subspace. This property is not restricted to small systems. 

Contrast to regular motions, where the trajectories are confined to
specific regions in the phase space, chaotic trajectories can travel
freely within the constant energy subspace of the phase
space. Therefore, chaotic motions are sometimes used for simulations
of finite temperature systems. Indeed, apart from quantum effects,
this should be able to describe a real physics system, given enough
degrees of freedom. The temperature for such systems can be well
defined, but it is unclear whether the properties of finite
temperature systems are reproduced, in general. In this work,
Hamiltonian systems are used to simulate the dynamics of finite
temperature lattice systems, and the physical properties of the system
is investigated from the point of view of the basic thermodynamic
laws. It is found that the basic law that the temperature within a
thermally equilibriated system is uniform, can be violated in certain
cases. We demonstrate this with a few examples, and analyze why it can
occur.

Chaoticity of a system can be analyzed quantitatively through its
Lyapunov exponents. For this purpose, the time averaged Lyapunov
exponents over trajectories are usually studied. For autonomous
Hamiltonian systems, these exponents are composed of pairs, which each
sum to zero. One can also study the local, or the finite time,
Lyapunov exponents, which have not been averaged over
time. Interestingly, the local Lyapunov exponents are in general not
paired, even in autonomous Hamiltonian systems, but seem to become
paired after some time. This pairing property is studied in this work
for the $\phi^4$ theory. It is found that the pairing time is faster
at higher energies, and that the symmetries of the system can affect
the pairing time, along with the convergence properties of the
Lyapunov exponents.
For concreteness and consistency, we shall work mostly with the one
dimensional lattice $\phi^4$ theory in this work, while explaining the
differences and similarities with other models, when appropriate.  

The symmetric properties of the dynamical system, and its effects of the chaotic
trajectories is analyzed in \sect{symmetry}. The chaotic properties of
the dynamical systems in relation to thermodynamics relations are studied in
\sect{temp}. The pairing properties of local Lyapunov exponents are
discussed in \sect{lyapunov}.
\section{Symmetries and chaos}
\label{sec:symmetry}
Dynamical systems, in general, have various symmetries that can constrain some
of its dynamics. These symmetries depend strongly on the structure of
the model considered.
For concreteness, we use the $\phi^4$ theory in one dimension with $N$
lattice sites\cite{AK1,Hu}. The Hamiltonian for the model is
\begin{equation}
    \label{eq:ham} H=\sum_{j=1}^N  {p_j^2\over2}
      +\sum_{j=1}^{N
      -1}{(q_{j+1}-q_j)^2\over2}+H_{\rm B}+\sum_{j=1}^N{q_j^4\over4}
    \quad,
\end{equation}
The potential terms at the ends, $H_{\rm B}$, depend on the boundary
conditions, and are 
\begin{equation}
    \label{eq:lb} 
    H_{\rm B}= {1\over2}(q_{N}-q_1)^2\quad\hbox{(periodic bc)},
    \quad H_{\rm B}= {1\over2}\left(q_{N}^2+q_1^2\right)
    \quad\hbox{(fixed bc)},\quad H_{\rm B}= 0\quad\hbox{(free bc)}.
\end{equation}
The non-linearity of the model is contained solely in the {\it local}
on-site potentials. The intersite couplings of $q_j$'s induce energy
transfer across the sites.  The model can be chaotic when $N\geq2$.
The equations of motion for the model are
\begin{equation}
    \label{eq:eom}
    \dot q_j =p_j,\quad     \dot p_j= q_{j+1}+q_{j-1}-2q_j -q_j^3
    \quad j= 1,2,\cdots,N\quad.
\end{equation}
with the boundary conditions, which can be interpreted as the
following, by adding a site at each end.
\begin{equation}
    \label{eq:eomBC}
    q_0=q_N,q_{N+1}=q_1\ \hbox{(periodic)},\quad
    q_0=q_{N+1}=0\ \hbox{(fixed)},\quad
    q_0=q_1, q_{N+1}=q_N\ \hbox{(free)}\quad.
\end{equation}
We can consider more exotic boundary conditions, $q_1=aq_{N+1}$, ($a$:
constant) which we shall call twisted boundary conditions. This
corresponds to adding the potential,
\begin{equation}
    \label{eq:twisted} 
    H_{\rm B}= {1\over2}(q_1-aq_{N})^2\quad\hbox{(twisted bc)}\quad.
\end{equation}
An useful example is the antiperiodic boundary condition,
$q_1=-q_{N+1}$, with the notation used in \eqnn{eomBC}. Antiperiodic
boundary condition can appear naturally within even within other
boundary conditions, as will be seen below. Antiperiodic boundary
conditions are also used in fermionic theories.

The minimal $\phi^4$ theory, explained above, has a ${\mathbb Z}_2$
symmetry, $(q_j,p_j)\leftrightarrow (-q_j,-p_j)$. The symmetry can be
enlarged by letting $(q_j,p_j)$ belong to a representation of a
group. The simplest example would be to let $(q_j,p_j)$ be complex,
and let the potential be a function of the complex norm of $q_j$. In
this case, there is an additional U(1) symmetry, which rotates the
phase of $(q_j,p_j)$. Such boundary conditions are used in the
theories of parastatistics\cite{para}, anyons\cite{anyon}, theories on
orbifolds\cite{orbifold}, and higher dimensional representations can
lead to to other interesting twisted boundary conditions\cite{thooft}.
Here, we shall {\em not} enlarge the symmetry, but work
with the simplest minimal $\phi^4$ theory.

In addition to the ${\mathbb Z}_2$ symmetry, there are other
spatial symmetries in the lattice model, which depend on the boundary
conditions. Here, we consider the model with the periodic boundary
conditions, so that the system is essentially on a ring.
 When the boundary condition is periodic,
for any factor $m$ of $N$, the system has a translational symmetry,
the symmetry under the transformation, ${\cal T}_m$, that shifts all
the sites by $m$,
$ {\cal T}_mq_j = q_{j+m}$.  This reduces the model to the $\phi^4$
theory with $m$ sites. When $m$ is not a factor of $N$, the symmetry
is incompatible with the boundary conditions.
Here, for the model with periodic boundary conditions, we use the convention
of identifying the site labeled by $j+N$ with that labeled by $j$.
There is another symmetry,
$(q_j,p_j)\leftrightarrow(q_{N-j},p_{N-j})$, $(j<N/2)$ which is
essentially the parity symmetry, for any $N$. This reduces the model
to that of $N/2+1$ sites when $N$ is even, and $(N+1)/2$ sites when odd.
When $N$ is even, there is also an inequivalent symmetry,
$(q_j,p_j)\leftrightarrow(q_{N+1-j},p_{N+1-j})$, $(j\leq N/2)$, which
effectively reduces the number of sites to $N/2$.  In fact,
$(q_j,p_j)\leftrightarrow(q_{M-j},p_{M-j})$, for any integer $M$ is a
symmetry of the model, for any $N$. However, by shifting the labels of
the sites, it reduces to the symmetries explained above.
 The model with other boundary conditions can also be considered, with their
corresponding symmetries. 

\begin{figure}[htbp]
  \centering
    \includegraphics[width=8.6cm,clip=true]{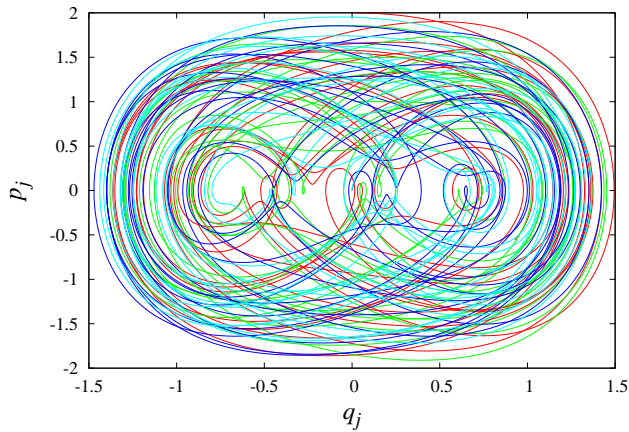}
    \caption{Chaotic trajectories, $(q_j,p_j)$ for the initial
      conditions, $q_j=0, (p_j)=(2,1,0,0,2,1,0,0,2,1,0,0)$ for $N=12$,
      from $t=0$ to $t=100$.  There are only four distinct
      trajectories due to symmetry, sites 1 (red), 2 (green), 3
      (blue), 4 (cyan).}
  \label{fig:trajSym}
\end{figure}
\begin{figure}[htbp]
  \centering
    \includegraphics[width=8.6cm,clip=true]{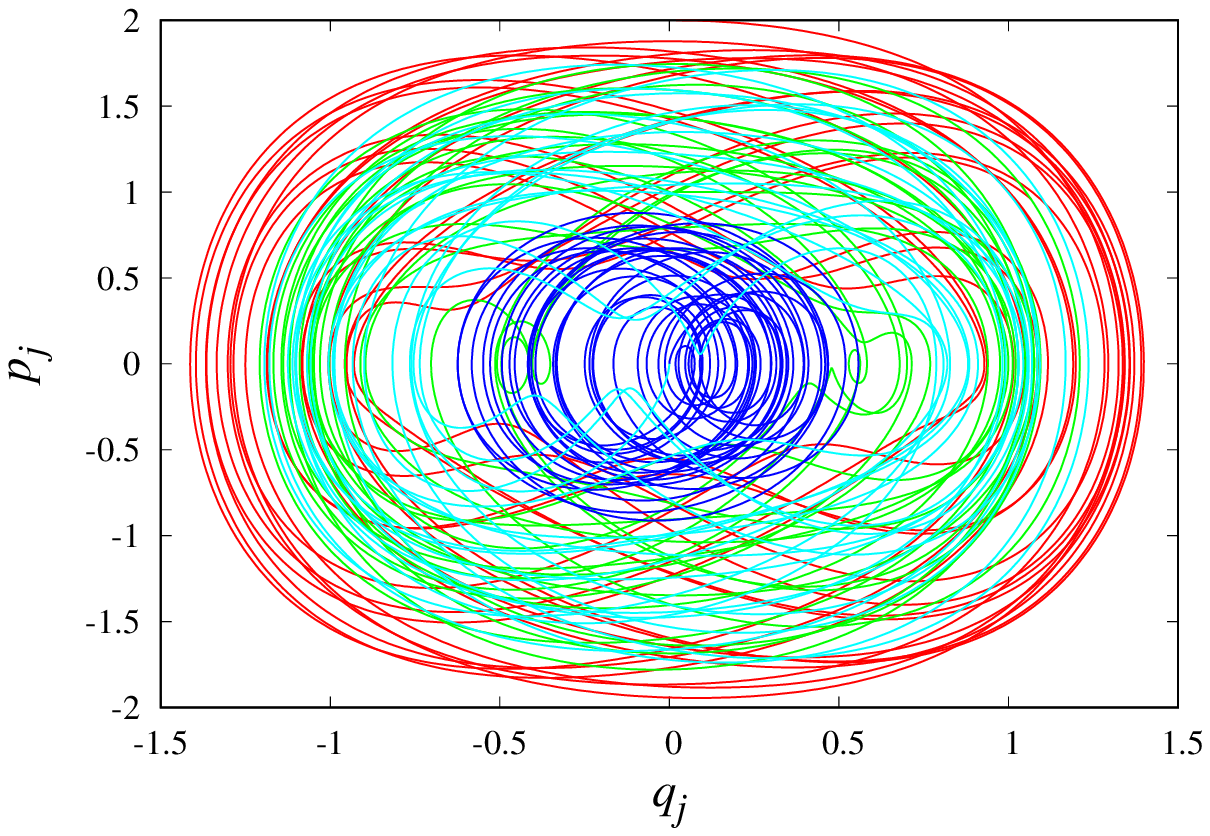}
    \includegraphics[width=8.6cm,clip=true]{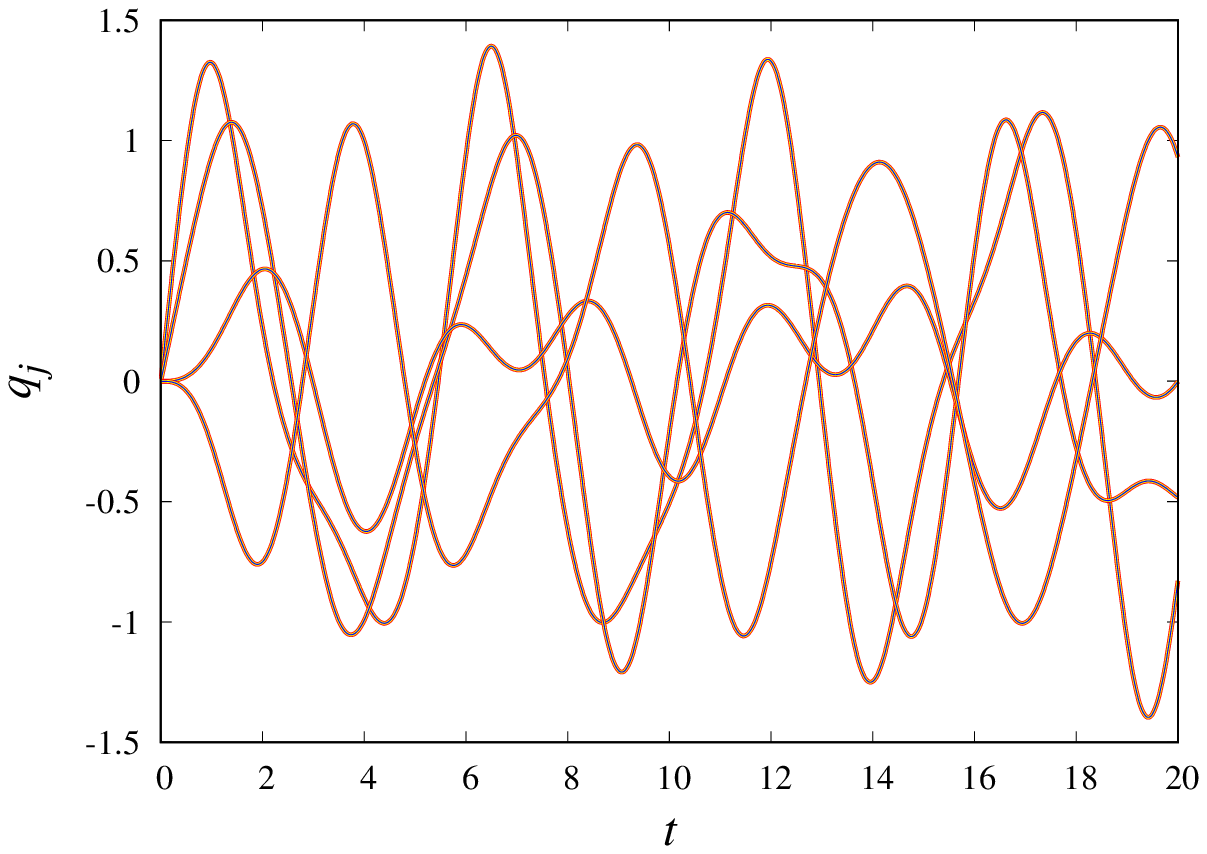}
    \caption{(Left) Chaotic trajectories, $(q_j,p_j)$ for the initial
      conditions, $q_j=0, (p_j)=(2,1,0,0,-2,-1,0,0)$ for $N=8$, from
      $t=0$ to $t=100$. There are only four distinct trajectories, up
      to sign, due to symmetry, sites 1 (red), 2 (green), 3 (blue), 4
      (cyan).  (Right) Coordinates $q_j$ (red), $ -q_{j+4}$ (yellow)
      as a function of time. Trajectories $q_j=0, (p_j)=(2,1,0,0)$ for
      $N=4$ with antiperiodic boundary conditions are also shown
      (blue).  The trajectories for $q_j,-q_{j+4}\ (N=8)$, and
      $q_j \ (N=4)$ are identical for $j=1,2,3,4$, so that the symmetry
      of the trajectory for the $N=8$ system can be observed, as well
      as the equivalence to the $N=4$ system with antiperiodic
      boundary conditions.}
  \label{fig:trajAsym}
\end{figure}
In theories with chaos,  generic trajectories thread through the
allowed region, and since the different trajectories generically
diverge from each other, it seems difficult to restrict the chaotic
trajectories within a subspace. Here, we use the $\phi^4$ theory
example to show how a symmetry of a model can restrict the
trajectories to a subspace of a dynamical system, while still being chaotic.
First, let us use the translational symmetry, ${\cal T}_m$, where $m$
is a factor of $N$: When the initial conditions also respect this
symmetry, the equations of motion reduce to the $\phi^4$ model on the
one dimensional lattice with $m$ sites. The solutions to the equations
of motion is restricted to the subspace,
$\{(q_j,p_j)|q_{j+m}=q_j,p_{j+m}=p_j, j=1,2,\ldots,N\}$, for {\em the
  initial conditions respecting this symmetry}. The equations of
motion can be seen to be consistent with the reduction to this
subspace. Here, we used the convention,
$q_{j+N}=q_j,p_{j+N}=p_j$. Excluding the trivial case $m=1$, the
motion within this subspace is in general chaotic, since it is
identical to that of the one-dimensional $\phi^4$ model for $m$ sites,
which can be chaotic, as mentioned previously.  An example of this
symmetry for $N=12$, $q_{j+4}=q_j$, $p_{j+4}=p_j$, is shown in
\figno{trajSym}. In this example, the maximal Lyapunov exponent,
$\lambda_1=0.03$. Lyapunov exponents are discussed systematically in
\sect{lyapunov}.

\begin{figure}[htbp]
  \centering
    \includegraphics[width=8.6cm,clip=true]{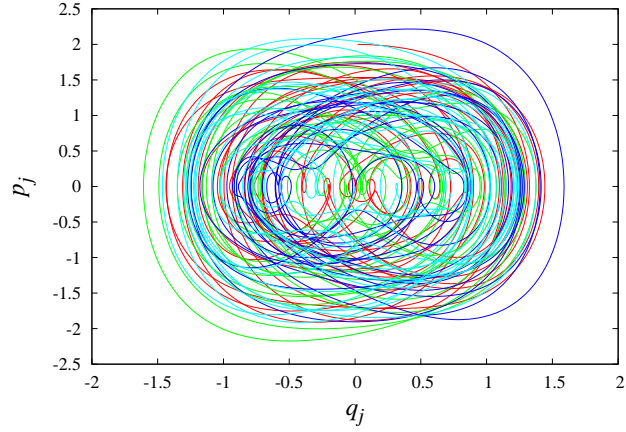}
    \caption{Chaotic trajectories, $(q_j,p_j)$ for the initial
      conditions, $q_j=0, (p_j)=(2,1,0,0,-2,-1,0,0,2,1,0,0)$ for
      $N=12$, from $t=0$ to $t=100$. The trajectories for first four
      sites are shown, sites 1 (red), 2 (green), 3 (blue), 4
      (cyan). }
  \label{fig:trajAsymNo1}
\end{figure}
\begin{figure}[htbp]
  \centering
    \includegraphics[width=8.6cm,clip=true]{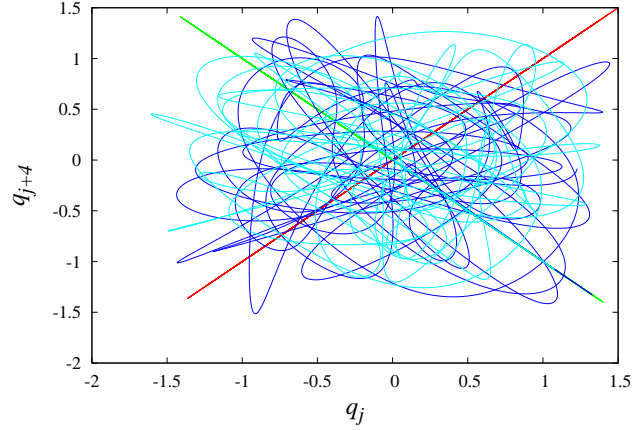}
    \caption{The phase space relations of $(q_1,q_5)$ for
      the trajectories shown in \figno{trajSym} (red), \figno{trajAsym}
      (green), and \figno{trajAsymNo1} (blue). $(q_2,q_6)$ trajectory for
      \figno{trajAsymNo1} is also shown (cyan).  We see that $q_1=q_5$ for the
      trajectory in \figno{trajSym}, $q_1=-q_5$ for that in
      \figno{trajAsym}, and that no such simple relation exists for
      $(q_1,q_5), (q_2,q_6)$, for the trajectory in \figno{trajAsymNo1}.}
  \label{fig:trajAsymNo2}
\end{figure}
A different, and slightly less obvious symmetry can be imposed when
$N/m$ is even, with $m$ being a non-trivial factor of $N$. Using the
${\mathbb Z}_2$ symmetry of the $\phi^4$ theory, we can restrict the
orbits to the subspace,
$\{(q_j,p_j)|q_{j+m}=-q_j,p_{j+m}=-p_j, j=1,2,\ldots,N\}$.  The
equations of motion are consistent with this restriction, provided
that $N/m$ is even.  In this case, the dynamics  of the  $m$ sites reduce to
the $\phi^4$ theory with {\em anti}periodic boundary conditions. In
\figno{trajAsym}, an example of this symmetry for the case $N=8,m=4$
is shown. For this trajectory, $\lambda_1=0.06$.  For comparison, in
\figno{trajAsymNo1}, an example for the $N=12$ case, with initial
conditions similar to those used for $N=8$, is shown, which has
$\lambda_1=0.05$.  This case breaks the symmetry, so that the
trajectories are no longer confined to a lower dimensional subspace
within the constant energy subspace.
The trajectories that do and do not respect symmetry properties are
contrasted in \figno{trajAsymNo2}.

\begin{figure}[htbp]
  \centering
    \includegraphics[width=8.6cm,clip=true]{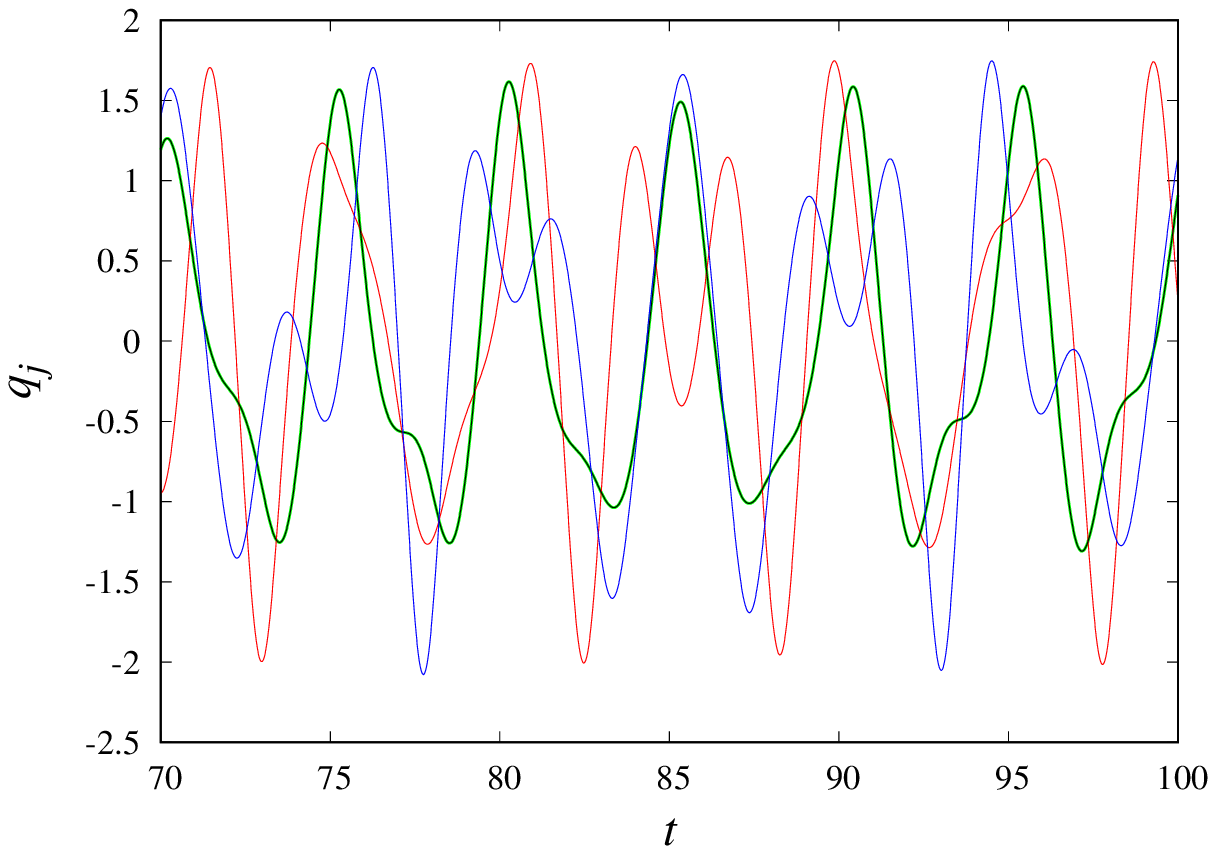}
    \includegraphics[width=8.6cm,clip=true]{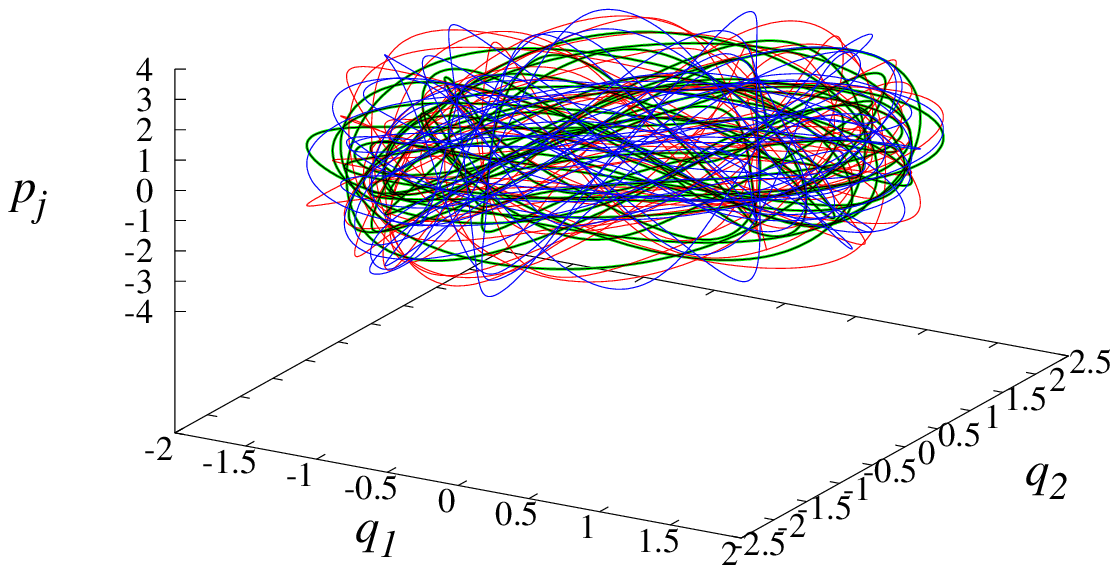}
    \caption{(Left) The time dependence of $q_j$ on time with the
      initial conditions, $q_j=0, (p_j)=(2,2,2,-2)$ for $N=4$; $q_1$
      (green), $q_2$ (red), $q_3$ (black), and $q_4$ (black).  It can
      be seen that $q_1=q_3$, and $q_1,q_2,q_4$ have no such simple
      relation. (Right) Chaotic trajectories of $p_j$ against
      $(q_1,q_2)$ from $t=0$ to $t=100$, with the same colors as the
      left figure.  }
  \label{fig:trajZ2}
\end{figure}
\begin{figure}[htbp]
  \centering
    \includegraphics[width=8.6cm,clip=true]{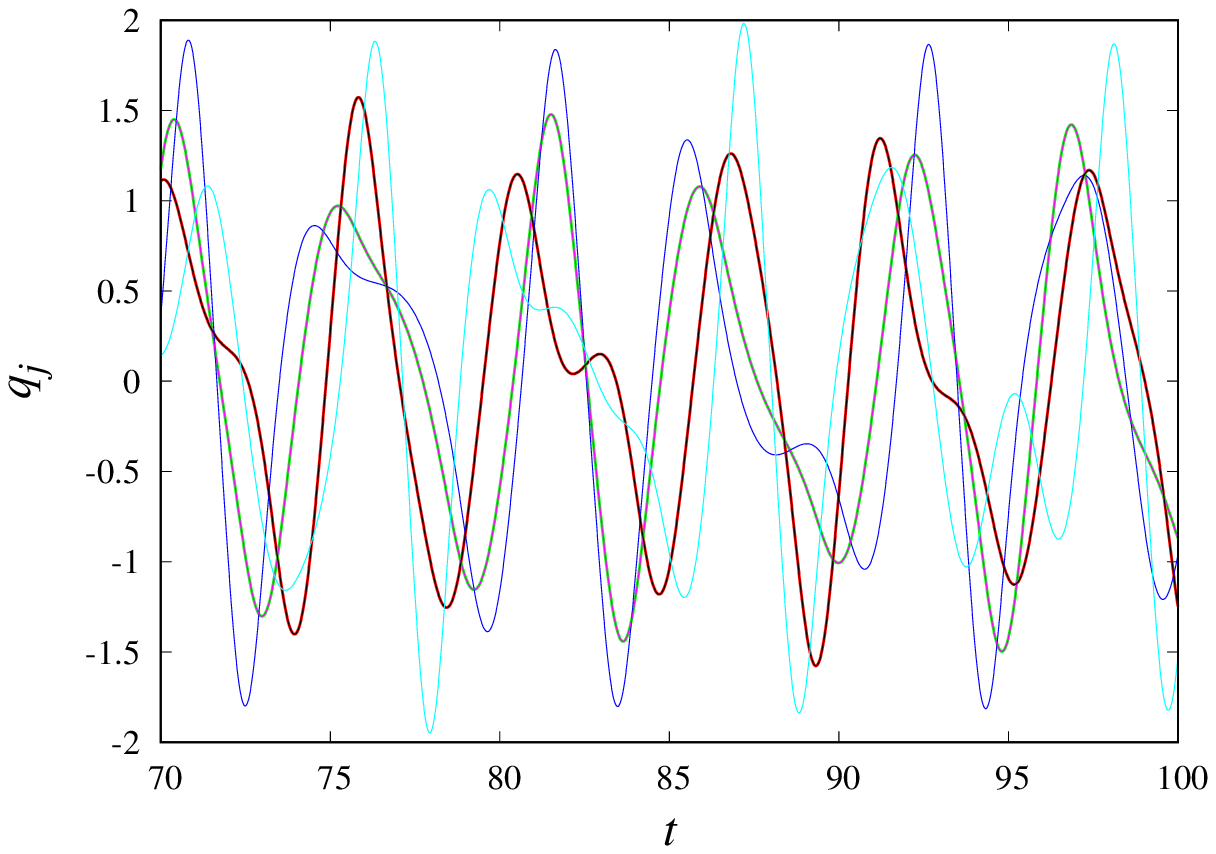}
    \includegraphics[width=8.6cm,clip=true]{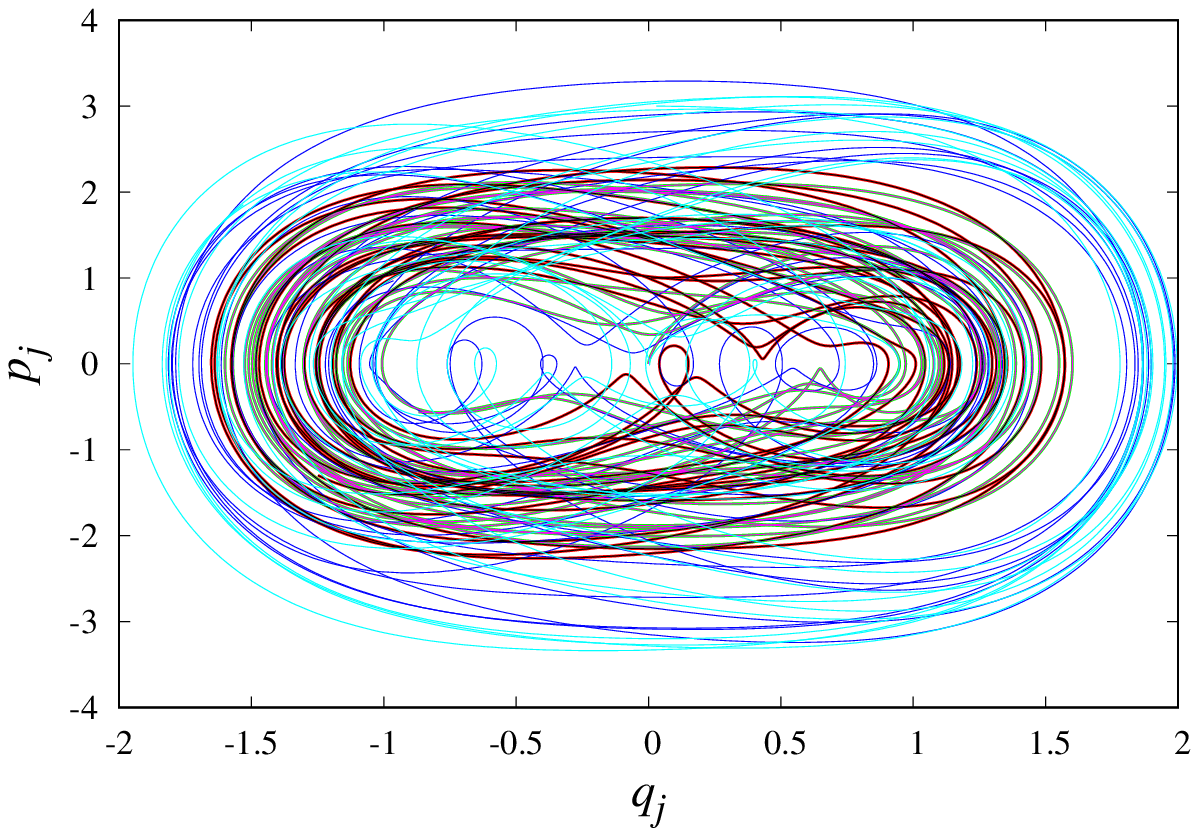}
    \caption{(Left) The time dependence of $q_j$ on time with the
      initial conditions, $q_j=0, (p_j)=(1,0,2,0,1,3)$ for $N=6$; $q_1$
      (red), $q_2$ (green), $q_3$ (blue), $q_4$ (magenta) $q_5$
      (black), and $q_6$ (cyan).  It can be seen that
      $q_1=q_5,q_2=q_4$, and there are no other simple
      relations. (Right) Chaotic trajectories of $p_j$ against
      $(q_1,q_2)$ from $t=0$ to $t=100$, with the same colors as the
      left figure.  }
  \label{fig:traj3}
\end{figure}
Another interesting, and somewhat obscure symmetry that can restrict the
chaotic trajectory to a subspace is the parity symmetry mentioned
above. Here, when $N$ is even, we can restrict the dynamics to a
subspace, $\{(q_j,p_j)|q_{j}=q_{N-j},p_{j}=p_{N-j}, j=1,2,\ldots,N\}$.
In this case, the phase space coordinates
$(q_{N},p_N), (q_{N/2},p_{N/2}) $ are not restricted at all. A simple
interesting example for $N=4$, and an example for $N=6$ are shown in
\figno{trajZ2}, and \figno{traj3}. $\lambda_1=0.02, 0.04$, for these
trajectories, respectively, and the chaoticity of the former
trajectory is revisited in \sect{lyapunov}.
It should be noted that in all the cases, the dynamics preserves the
symmetries of the model, only provided that the initial conditions
respect them.

Above, the explicit examples used to illustrate the analysis were
relatively small systems ($N\leq12$). However, clearly, the symmetry
properties apply to theories with arbitrarily large lattice sizes,
$N$. For instance, for an arbitrarily large $N$, a translational
symmetry ${\cal T}_m$ with $m$ being any factor of $N$ (which can also
be large) exists. In this case, the non-trivial dynamics of the system
reduces to that of the $\phi^4$ theory with $m$ lattice sites. This
also almost guarantees that the motion restricted to a lower
dimensional subspace within the phase space can be chaotic, since it
is the dynamics of the $\phi^4$ theory which can have an arbitrarily
large number of sites.
While we adopted the $\phi^4$ model for concreteness, it should be
noted that this symmetry argument applies straightforwardly to a
one-dimensional lattice models with any {\em on-site } potential, that
is even under the reflection $q_j\leftrightarrow -q_j$. The
translational symmetry can still be used even when the potential is
not even. The symmetry can also be generalized to higher dimensional
lattice theories.
\section{Chaos, ideal gas temperature,  and thermodynamic laws}
\label{sec:temp}
In theories governed by autonomous Hamiltonians, if we follow
trajectories over time, they are constrained in a constant energy
subspace. If the dynamical system has chaos, we might wonder if it goes
``everywhere'' within this subspace, or travel densely within
it. Clearly, there are cases where the motion is restricted: First,
there is a trivial possibility that the system might be decoupled into
subsystems with no interaction amongst them. In this case, there might
be chaos within each dynamical system, but  trajectories are restricted within
product spaces. These cases include systems which might seem coupled,
but whose phase space coordinates can be decoupled by canonical
transformations. Second, even in theories with chaos, there are
non-chaotic orbits, such as periodic
orbits\cite{Rosenberg,SanduskyPage,Flach,NLNMRev,ChechinRev,P4}. Then,
as we have seen above, symmetries can constrain a trajectory to a
subspace.  Relatedly, conservation laws, often associated with
symmetries, can also constrain the dynamics.
In a dynamical system not reducible to decoupled subsystems, when we
consider a initial condition that respect no symmetries, a question
naturally arises as to whether the trajectory ultimately travels
densely, or arbitrarily close to any point, within the constant energy
subspace. If such is the case, there is an unique chaotic ``sea'', and
the system is ergodic. Then, averaging over a chaotic trajectory
within the sea, we may obtain the unique statistical average of any
physical quantity. In an microcanonical average, the probability
distribution within the phase space is uniform over the constant
energy surface.
However, it should be noted that even in this idealized case, the time
required to sample broadly enough within the phase space to evaluate
the physical quantity might be prohibitively long, for computational
purposes. Also, even if the trajectory travels densely within the
constant energy subspace, the corresponding probability distribution
might not be uniform, so that the averaging is not microcanonical.

We now investigate these issues in conjunction with the notion of
temperature in Hamiltonian dynamics. While we will not reach a simple
conclusion, we obtain results that are interesting and deserve further
study. In classical theories, it is possible to theoretically apply
Nos\'e--Hoover thermostats and analyze its dynamics from first
principles\cite{thermo1,thermo2,HH1}. 
Deterministic thermostats, such as Nos\'e-Hoover thermostats, add
additional degrees of freedom to the system, that are not described by
Hamiltonian dynamics. When the equations of motion is integrated, the
thermostats induce thermal distributions for the degrees of freedom
coupled to the thermostats, as an ensemble when the trajectory is
sampled over time.
The temperature, in this context, can be measured by the ideal gas
temperature, which is defined as $\vev{p_j^2}$ for the site $j$ in the
lattice model, with $\vev{\cdots}$ denoting the time averaged value
over the trajectories, which is also the statistical ensemble average
when the system is ergodic.  This ideal gas temperature is
identical to any temperature definition, provided the site is
thermalized. In Hamiltonian dynamics, with{\it out} thermostats, one
can similarly define the ideal gas temperature locally, for any
site. We shall use this definition of the temperature in this work. 
A natural question remains as to whether its behavior is such
that it can be interpreted as a thermodynamical temperature.
It should be mentioned that a combination of the coordinates, $q_j$,
may also be considered the definition of the temperature. However, in
lattice models such as the $\phi^4$ theory, $q_j$ are coupled across
sites, and it has a non-linear potential. For these reasons it
difficult to use coordinates in a simple unambiguous definition of the
local temperature.

In this work, we consider closed autonomous lattice Hamiltonian
systems.  There are no thermostats in the system, and naively, the
``temperature'' of all the sites should be identical, in the steady
state. If thermodynamic laws apply, the temperature of any matter
that is able to exchange energy with each other would be all at the same 
temperature, in a thermal equilibrium.
One objection might be that thermodynamics only applies to systems
with many degrees of freedom and small systems need not satisfy the
law. However, when the averaging is performed over large number of
configurations, typically by integrating over a trajectory, the local
temperatures are well defined and the ensemble average should satisfy
the thermodynamic laws.  In fact, in thermostatted systems,
thermodynamic laws apply well to systems with small degrees of
freedom\cite{HH1}. Another issue, which is more essential here, is the
assumption of thermalization. 
Here, thermalization means that the basic statistical properties of
the finite temperature systems are obeyed, so that the thermodynamics
laws apply\cite{thermo}.
If the system is not thermalized, while the ideal gas temperature
itself is well defined, the definition of the temperature is no longer
unique\cite{AK2}. For instance, we can define the temperature as
$\vev{p_j^2}$ or as $\sqrt{\vev{p_j^4}/3}$, and these values are in
general not equal. If the system is thermalized, the canonical
distribution $\exp(-H/T)$ dictates a Gaussian distribution for
$\{p_j\}$ since the momenta are quadratic and decoupled amongst the
sites in $H$, \eqnn{ham}. Therefore, using any even moment of $p_j$
leads to the same temperature.
When the system is large, the system is expected to be essentially
thermalized, after a sufficient time. After all, when the quantum
behavior is not essential, physical systems are governed by classical
dynamics and respect the thermodynamic laws. The microcanonical,
constant energy dynamics for the whole system should result in the
canonical ensemble for its subsystems.  However, for small systems,
there is no rigorous reasoning that the system should be
thermalized. It should be noted that even if the ideal gas
temperatures are different, it does not mean there is a way to extract
energy in a manner that violates the second law of thermodynamics,
since the law does not apply without thermalization.

The above logic leaves the possibility that in small Hamiltonian
systems, the ideal gas temperature is not identical within the whole
system, even with chaotic dynamics, in principle. However, even if
this is logically possible, it still remains to find examples of such
behavior, if it exists.  To our knowledge, no such system has been
found to date.  Here, we investigate this issue explicitly in the
$\phi^4$ model. First, we note that the ideal gas temperatures might
be identical due to symmetry reasons. Such is the case for the
$\phi^4$ model with {\em periodic} boundary conditions. The ideal gas
temperature is
\begin{equation}
  \label{eq:tempDef}
  \vev{p_j^2}=\sum_{\cal S}p_j^2
\end{equation}
where $\cal S$ is the chaotic sea for the given energy, over which it
is averaged.  By using the symmetry transformation in the previous
section, we find
\begin{equation}
  \label{eq:tShift}
  \vev{p_j^2}=\sum_{{\cal T}_1\cal S}{\cal T}_1 p_j^2=
  \sum_{\cal S}p_{j+1}^2  \quad.
\end{equation}
The first equality is simply the fact that shifting the site indices
is equivalent to relabeling sites. The second equality holds due to
the property ${\cal T}_m {\cal S}={\cal S}$, for any $m$ which is a
factor of $N$, which always includes the case $m=1$ used above. 
Given any trajectory $\{q_j\}$, $\{q_{j+m}\}$, which is
the solution just shifted by $m$, is also a solution, which leads to
this relation. 
This relation shows that {\it all} sites have the same ideal gas temperature. 
This assumes the uniqueness of the chaotic sea, for generic initial
conditions. One should be careful even in this case, since for a
particular {\em non-}generic initial condition, ideal gas temperatures
for the sites need not be the same. For instance, periodic orbits,
which are non-generic and clearly non-chaotic exist. In some cases,
some sites are stationary so that in this case, some sites are at zero
ideal gas temperature, while others are not\cite{P4}.  Also, as
studied in the previous section, we can construct dynamics with
chaotic behavior, yet restricted within a subspace.

\begin{figure}[htbp]
  \centering
    \includegraphics[width=8.6cm,clip=true]{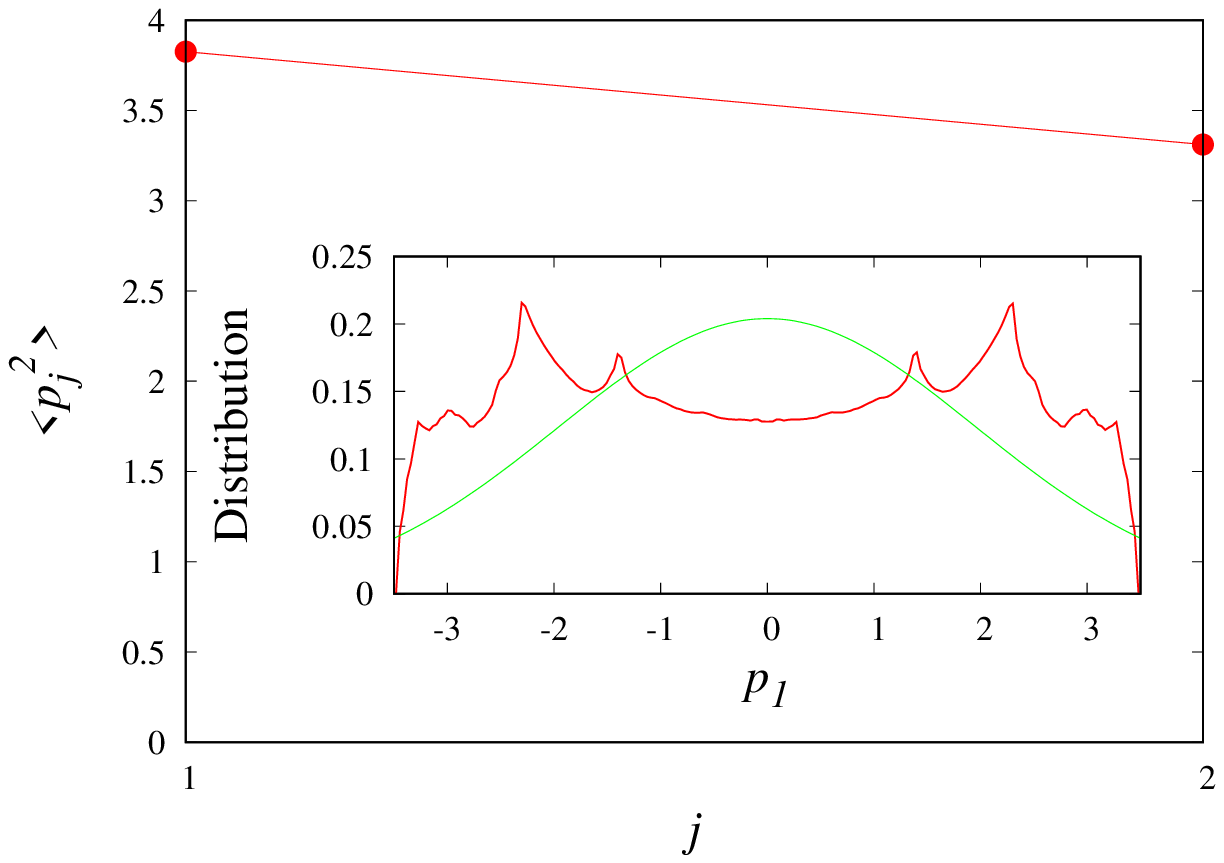}
    \includegraphics[width=8.6cm,clip=true]{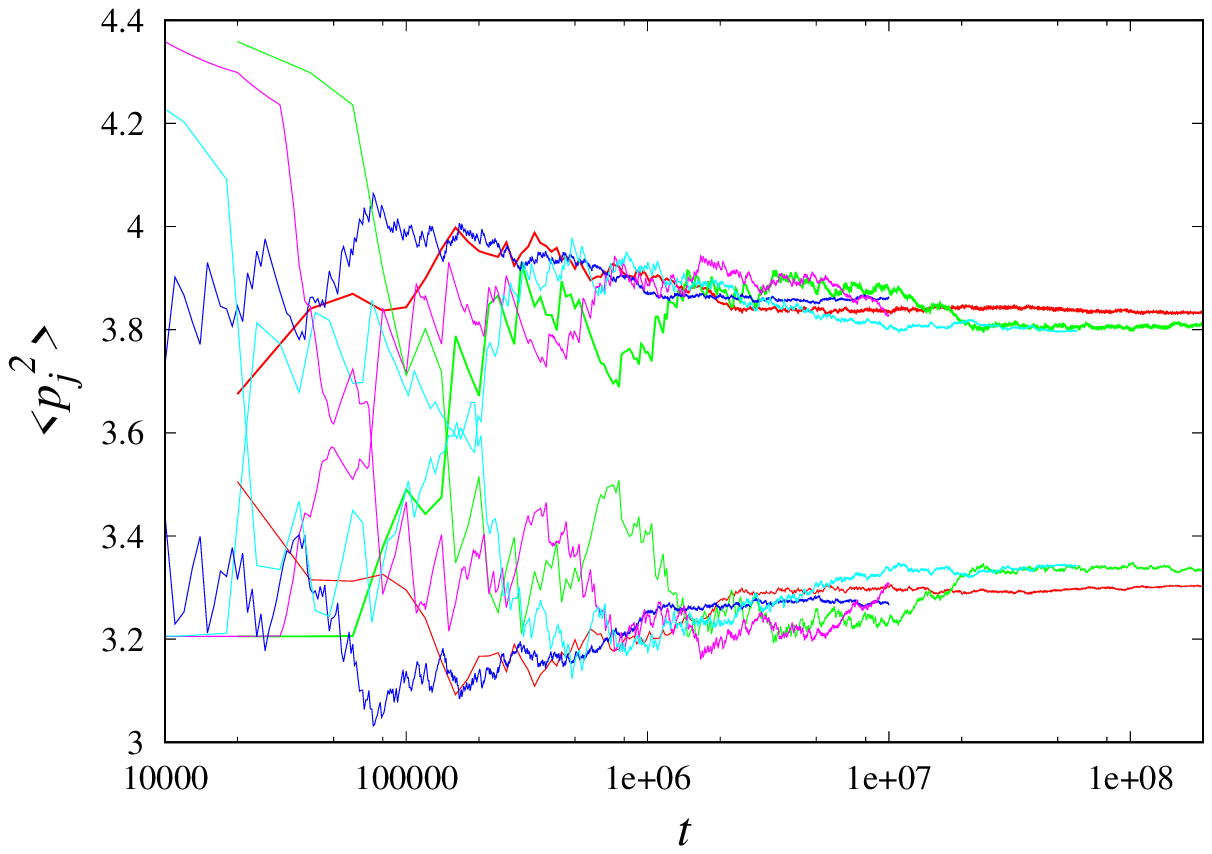}
    \caption{(Left) Ideal gas temperatures for $N=2$ $\phi^4$ theory
      with fixed, and free boundary conditions on the both ends. (Inset)
      $p_1$ distribution for site 1 (red), with the thermal
      distribution when $\vev{p_1^2}$ is regarded as the
      temperature (green). $(\nSample,\nStep,dt)=(10^4,2\times10^6,10^{-3})$. 
      (Right) The change in $\vev{p_j^2}$ with respect to
      the simulation time. It is seen that their averages are quite
      stable, and agree for various simulation parameters.  Parameters
      were $(\nSample,\nStep,dt)=(10^5,2\times10^6,10^{-3})$ (red),
      $(10^4,2\times10^7,10^{-3})$ (green),
      $(10^4,2\times10^6,5\times10^{-4})$ (blue),
      $(10^3,2\times10^7,5\times10^{-4})$ (magenta),
      $(10^4,2\times10^6,3\times10^{-3})$ (cyan).
}
  \label{fig:N2}
\end{figure}
\begin{figure}[htbp]
  \centering
    \includegraphics[width=8.6cm,clip=true]{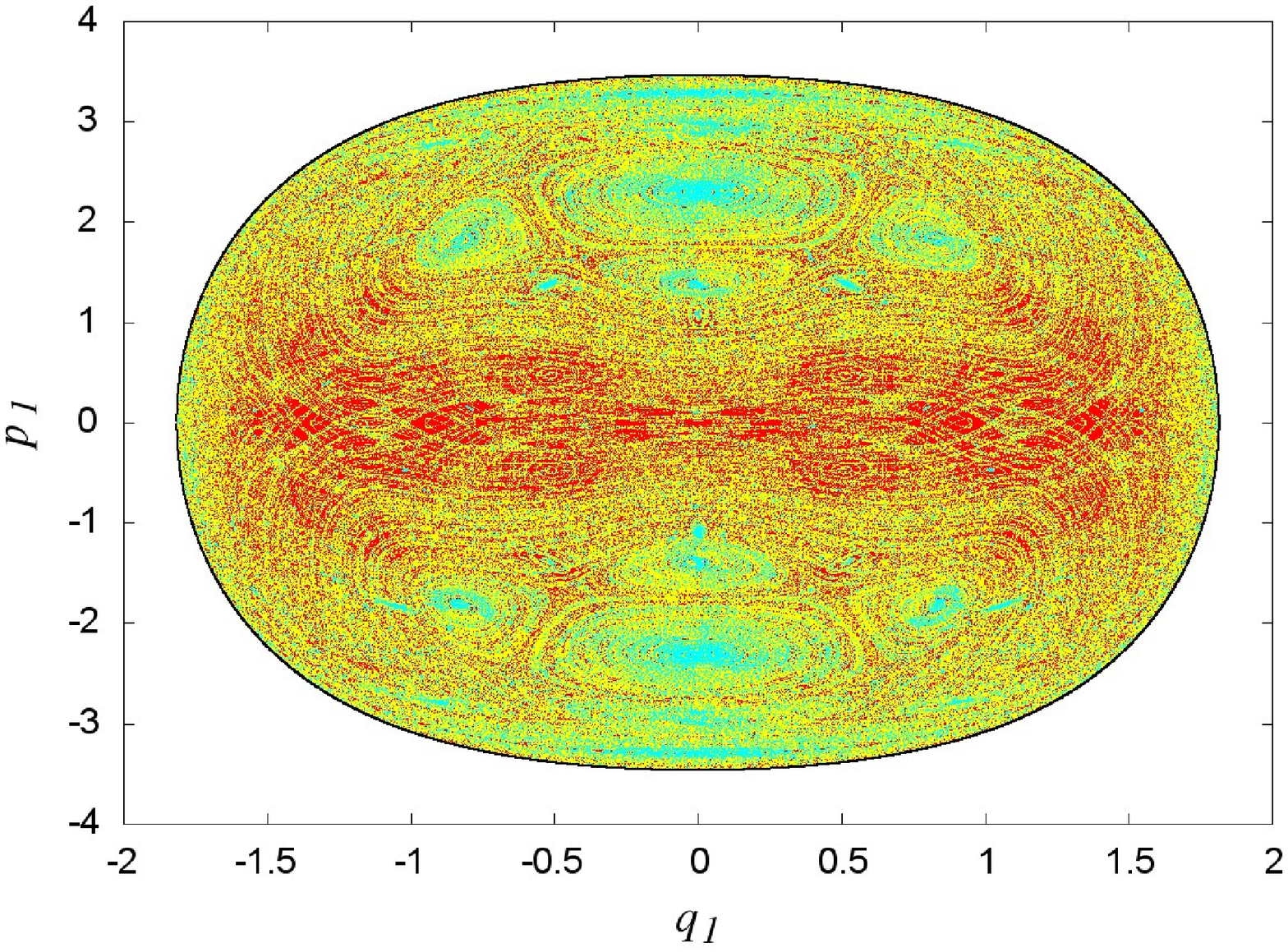}
    \includegraphics[width=8.6cm,clip=true]{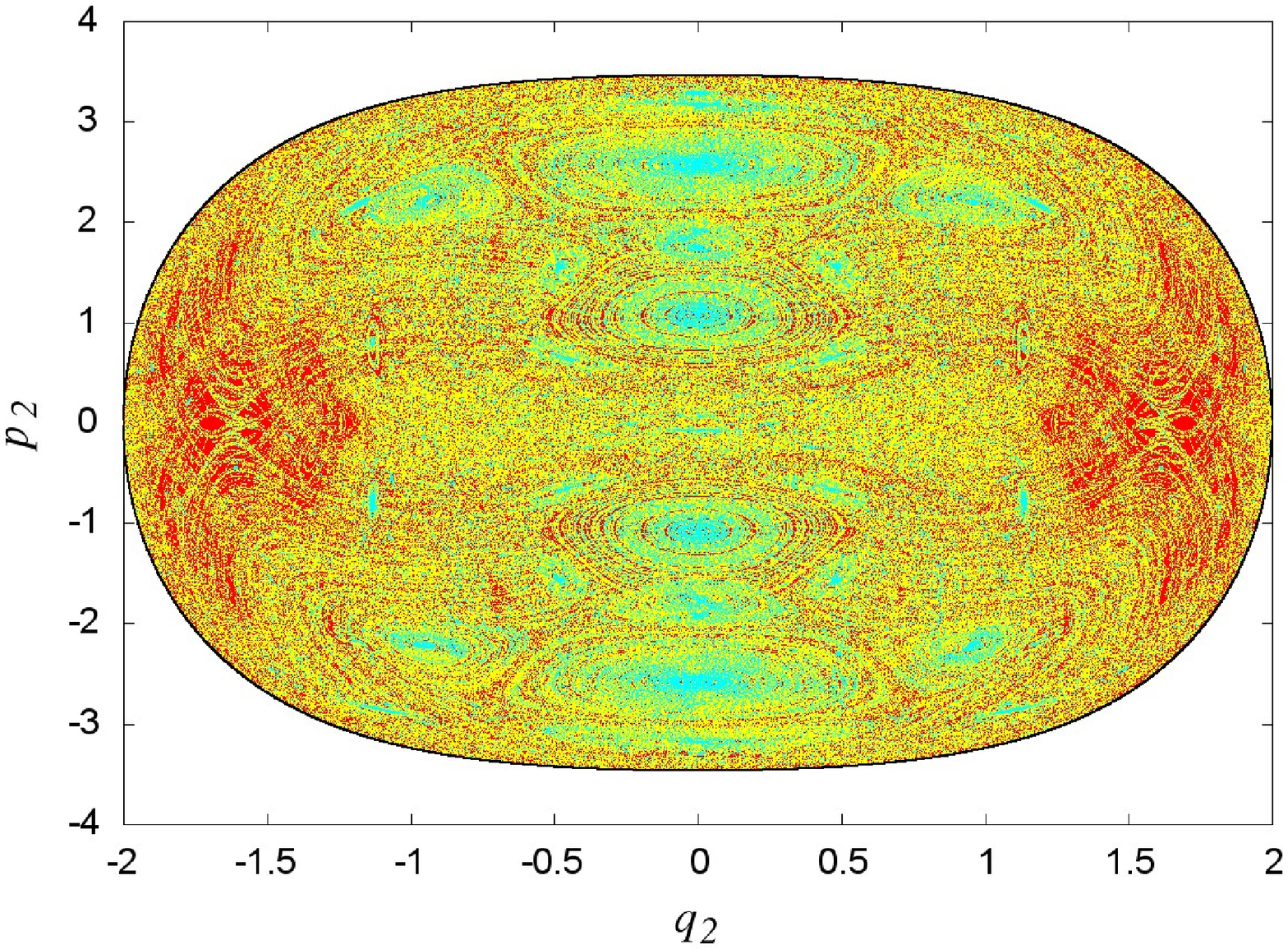}
    \caption{Poincar\'e sections for $(q_1,p_j)$ (left), $(q_2,p_2)$
      (right). The sections for the simulations with
      $(\nSample,\nStep,dt)=(10^3,2\times10^5,10^{-3})$ (cyan),
      $(\nSample,\nStep,dt)=(10^3,2\times10^6,10^{-3})$ (yellow),
      $(\nSample,\nStep,dt)=(10^5,2\times10^6,10^{-3})$ (red).  For
      the two sections, the number of points are
      $(103390,117830), (1038649,1179563),(104051927,117436859)$,
      respectively for the above three simulations. The boundary of
      the energetically is also shown (black).  The Poincar\'e
      sections fill out the allowed regions, as far as it can be
      observed.  }
  \label{fig:N22PS}
\end{figure}
\begin{figure}[htbp]
  \centering
    \includegraphics[width=8.6cm,clip=true]{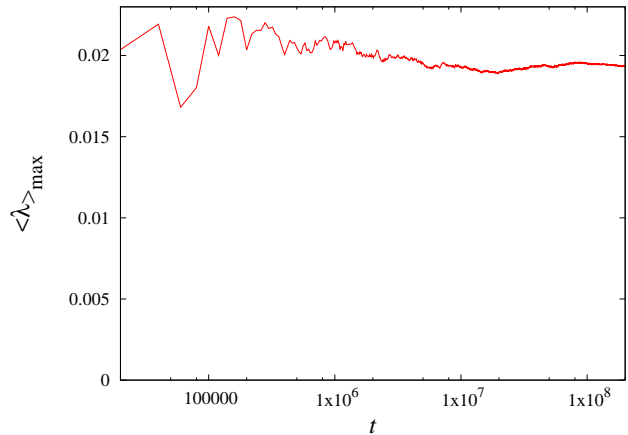}  
    \caption{The dependence of the averaged maximum Lyapunov exponent
      on the simulation time.
      $(\nSample,\nStep,dt)=(10^5,2\times10^6,10^{-3})$.  }
  \label{fig:N22Lyap}
\end{figure}
A concrete, and  interesting question is whether, in a microcanonical
ensemble average of a lattice Hamiltonian system, all the sites
have the same ideal gas temperature. Here, we provide evidence of a
few examples where the temperatures are {\it not} the same. Ideally we would
like to find the simplest examples, with generic parameters. 
Let us first consider the $N=2$ model. Here, the symmetry would
prevent us from having different temperatures in the two sites, if the
boundary conditions are identical, so we choose the model with the
fixed and free boundary conditions on the two ends. We pick $E/N=3$,
so that the parameters are of order one. Here, $E$ is the total energy
of the system, which is the value of the Hamiltonian, \eqnn{ham}.
This is, of course, constant along the trajectory, in theory. In
practice, the relative error in $E$ when integrated along the whole
trajectory is less than $10^{-10}$, relatively, in this work.  The
ensemble averaged ideal gas temperatures are shown in \figno{N2},
along with the change in their values over the simulation times. The
simulations are performed with randomized initial conditions with the
same energies, and averaging over the trajectories, and over the
initial conditions.  We see that $\vev{p_j^2}$ are unequal for the two
sites, considering the statistical errors. We add that when the
boundary conditions are the same, periodic, fixed, or free, at both
ends, the ideal gas temperatures are identical at both ends. The
momentum distribution for site $j=1$ is also shown in \figno{N2},
which can be seen to be far from the Gaussian distribution, which is
required for a thermalized system. To investigate as much as possible
that we have sampled in the constant energy subspace, we have changed
the lengths of each trajectory (number of steps $\nStep$), time step
size, $dt$, number of samples, $\nSample$, and also have performed the
computation with the fixed and free boundary conditions
reversed. Also, we have studied the trajectories within the phase
space, which seem to basically fill out the allowed region.
In \figno{N22PS}, the Poincar\'e sections in the $(q_1,p_1)$ plane
when $q_2=0$, and $(q_2,p_2)$ plane when $q_1=0$ are shown.
In the simulations, we found that when a single trajectory is used, it
tends to visibly leave regions of the phase space unvisited. On the
other hand, for simulations which seem to fill out the phase space,
the temperature profiles are consistent with the results obtained
above.  We have studied the projections of the trajectories on the
other coordinate planes, which also seem to fill out the allowed
region. To study if the trajectories are chaotic, as we expect, we
have computed the averaged maximal Lyapunov exponent, whose time
dependence is shown in \figno{N22Lyap}. This is seen to be stable and
non-zero.
\begin{figure}[htbp]
  \centering
    \includegraphics[width=8.6cm,clip=true]{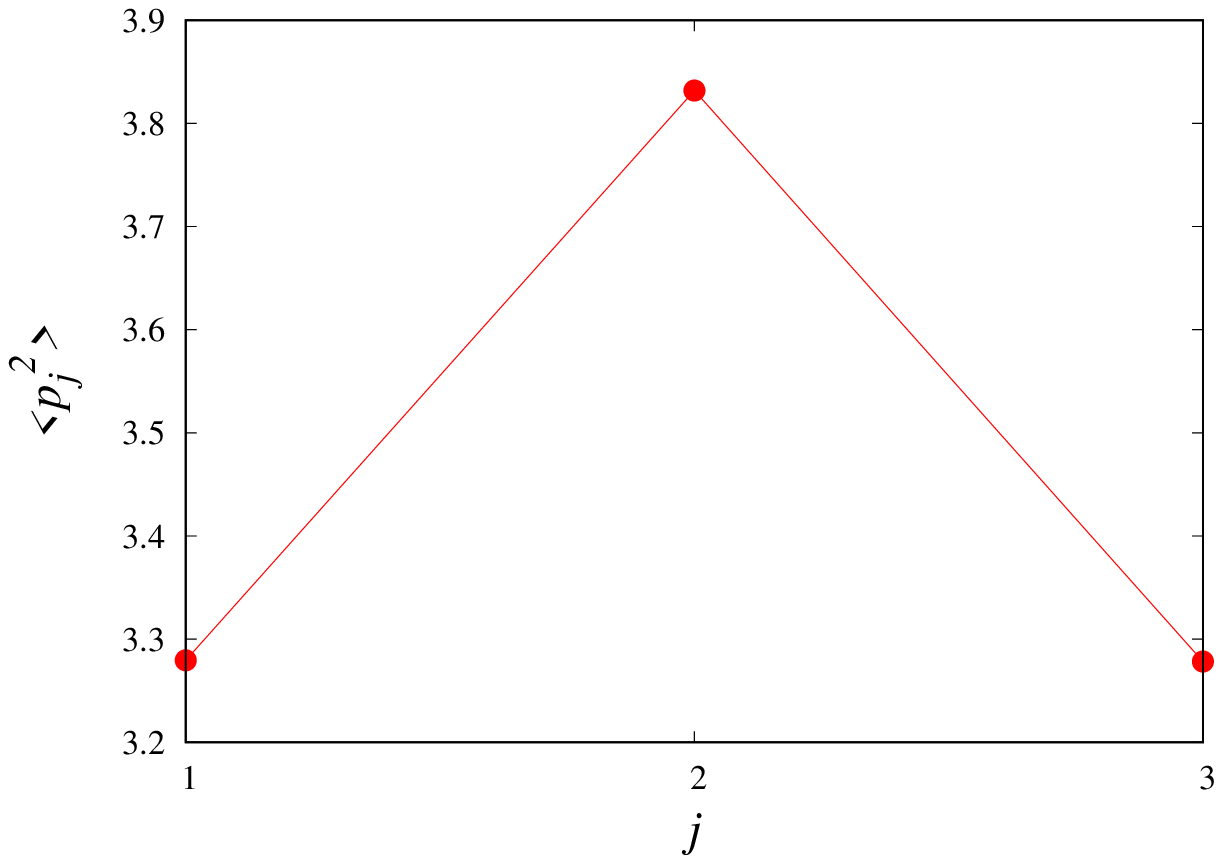}  
    \includegraphics[width=8.6cm,clip=true]{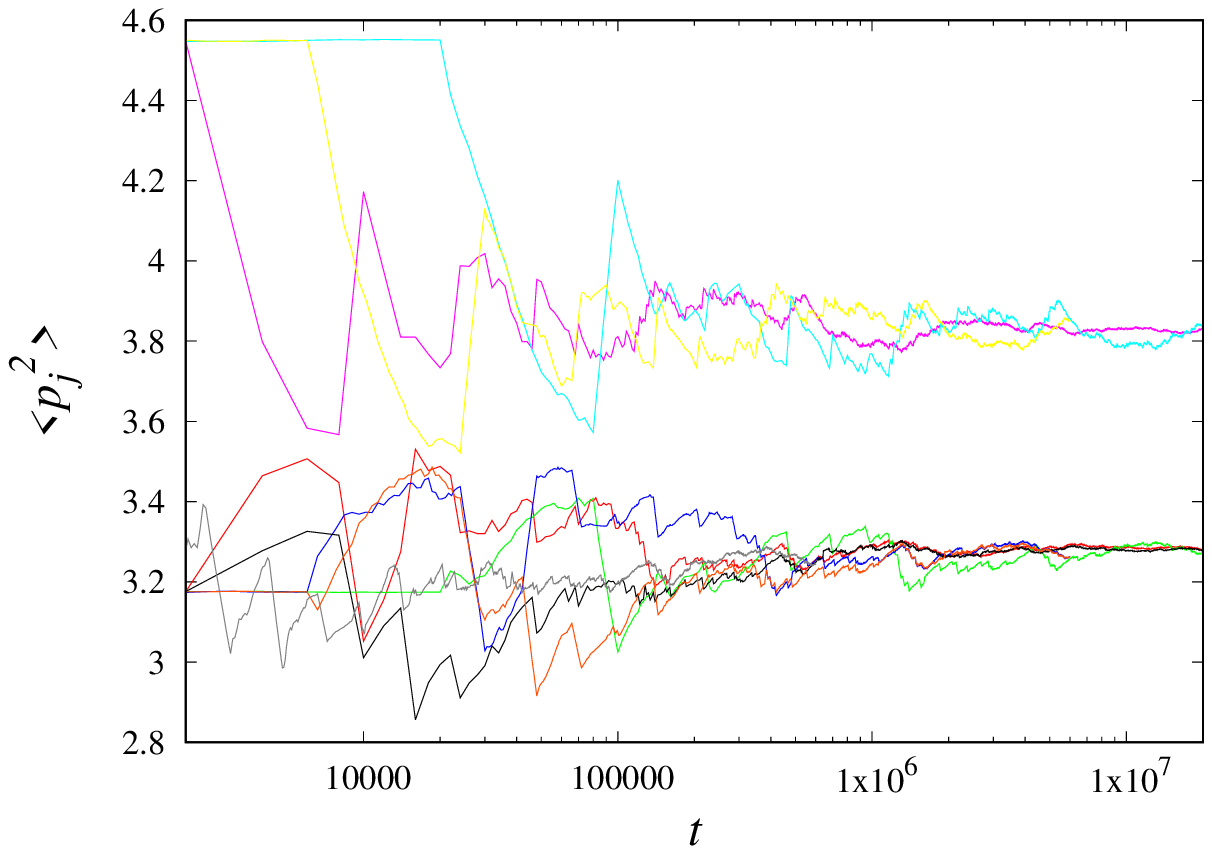}  
    \caption{(Left) Temperature profile for the lattice $N=3$ with
      fixed boundary conditions, for
      $(\nSample,\nStep,dt)=(10^4,2\times10^6,10^{-3})$.  (Right) The
      change in $\vev{p_j^2}$ with respect to the simulation time. It
      is seen that their averages are quite stable, and agree for
      various simulation parameters.  Parameters were
      $(\nSample,\nStep,dt)=(10^4,2\times10^6,10^{-3})$ (red, blue,
      green), $(10^3,2\times10^7,10^{-3})$ (magenta, cyan, yellow),
      $(10^3,2\times10^7,3\times10^{-4})$ (black, orange, gray), for
      the three sites in each parameter set. The first parameter set
      was used in the left figure. }
  \label{fig:N3}
  \end{figure}

  Next, we consider the $N=3$ model. In this case, as long as the
  boundary conditions are {\em not} periodic, there is no symmetry
  between the middle site and the sites at the ends, and we choose
  fixed boundary conditions at both ends. As above, the temperature
  profile and the dependence of the local temperatures on the
  simulation time, as well as the momentum distribution at site $j=1$
  is shown. \idt{1,3} at both ends agree, as they should, due to the
  symmetry of the Hamiltonian,
  $(q_1,p_1)\leftrightarrow(q_3,p_3)$. However, \idt{2} is not
  governed by symmetry, so it can be different, and is. When the
  boundary conditions are periodic, the ideal gas temperatures for all
  the sites are the same, as they should be.
In the above simulations, we used fourth order Runge-Kutta
algorithm\cite{NR} to integrate, with random initial conditions
generated for $(p_j)$ for a given total energy, $E$.  Mersenne
twister\cite{Boost} was mainly used for random number generation, with
some simulations using Knuth's pseudo random number generation
algorithm\cite{NR}.

We have argued why thermodynamic laws do not necessarily preclude
different ideal gas temperatures, in a simulation of a Hamiltonian
system, due to the lack of thermalization. We have performed some
simulations, which suggests that this indeed does occur in some
instances. However, this is not the last word on this interesting
subject, and the subject should studied more deeply to establish
whether such a difference indeed persists.
While a mathematical proof of the uniqueness of the chaotic sea is
undoubtedly difficult, we can examine its validity within this
example.  Here, we performed simulations within a set of initial
conditions, which seem to lead to a consistent result. This is
consistent with an unique chaotic sea.  However, for the single
trajectories that were studied, obvious lacunae in the phase space
often remained, within the limited simulation times. Whether more
simulation time will change this situation, within a realistic
simulation time, needs investigation.
\section{Pairing of local Lyapunov exponents}
\label{sec:lyapunov}
The chaotic properties of dynamical systems can be characterized by their
Lyapunov exponents, which shows how the neighboring trajectories
diverge exponentially from each other. Lyapunov exponents are obtained
by averaging local (or finite time) Lyapunov exponents over the phase
space trajectories\cite{Lyap0,Lyap1,Lyap2,chaos0,chaos1,HH1}. In
autonomous Hamiltonian systems, the (averaged) Lyapunov exponents are
paired in sets of $\pm\vev{\lambda}$, due to the time-reversal
symmetry. Furthermore, due to energy conservation, and this pairing
property, there is always at least a pair of zero exponents.

The local Lyapunov exponents are not necessarily paired, but when a
trajectory is followed in phase space, the exponents become paired, in
sets of two exponents with sum zero, after some
time\cite{HooverSnook,HH1}. We call this time the pairing time,
below. Since different exponents become paired at different times, the
pairing time refers to the time when all the exponents are paired. We
investigate what controls the pairing time below.  The pairing time
will depend on the dynamics, so a specific model is required for
comparison, for which we use the $\phi^4$ theory, to make use of some
of the results in the previous sections. For this study, we choose
$N=4$ with periodic boundary conditions, to elucidate some of the
properties pointed out in \cite{HooverSnook}. First, we expect the
pairing time depend on the physics parameters, such as the energy of
the system, and the initial conditions.  There are more technical
aspects, such as the type of integrator used, the ordering of the
coordinates, which hopefully do not play an essential role.

The pairing time clearly depends on the initial conditions, even at
the same energy, for the following reason. Since the local Lyapunov
exponents become paired after some time, the coordinates, including
the tangent vectors at that instant can be used as initial conditions,
which essentially means that the pairing time can be made small as
desired. For practical considerations, however, this is not useful,
since
finding these asymptotic coordinates and vectors itself requires
computation. So a more realistic problem is to start from a set of
initial coordinates, without requiring the integration of the
equations of motion, and to measure the pairing time.

\begin{figure}[htbp]
  \centering
  \includegraphics[width=8.6cm,clip=true]{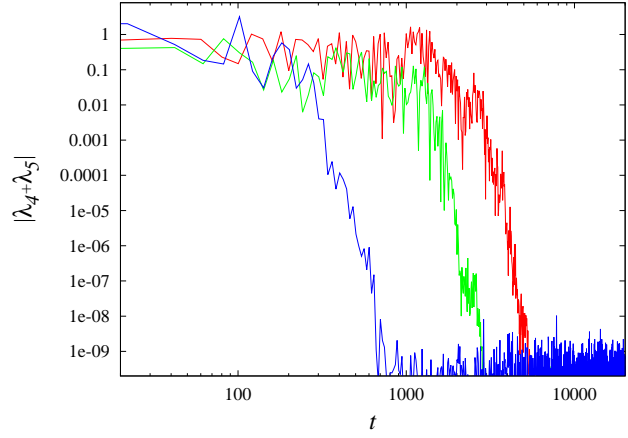}  
  \caption{Pairing behaviors, $|\lambda_4+\lambda_5|$, for
    $N=4\ \phi^4$ theory with respect to time, for typical initial
    conditions with $E/N=1$ (red), 2 (green), 10 (blue).  The pair corresponds to
    the exponents with the smallest averaged absolute values. The
    pairing occurs quite clearly, and $\tPair$ is smaller for larger
    $E/N$.  }
  \label{fig:pairE}
\end{figure}

\begin{table}[htbp]
  \centering
  \begin{tabular}{r|r|r|r}
    $E/N$&$\tPair<10^3$&$10^3<\tPair<10^4$&$\tPair>10^4$\\
    \hline
    0.1&0&0&1\\
    1&0&0.45&0.55\\
    2&0&0.8&0.2\\
    10&0.8&0.1&0.1\\\hline    
  \end{tabular}
  \caption{Dependence of $\tPair$
    on $E/N$, for the $N=4$ $\phi^4$ theory  with periodic
    boundary conditions. Ratio of the number of initial conditions
    for given $\tPair$ ranges,  out of 20 randomly generated initial conditions.
  } 
  \label{tab:pair}
\end{table}
Before we investigate the issue of initial conditions more thoroughly,
let us look at the dependence on energy. Intuitively, we expect that
larger total energies ($E$) lead to shorter time scales and shorter
pairing time.  This, however, is not obvious, since the oscillatory
frequency of the quadratic part of the Hamiltonian, \eqnn{ham}, does
{\em not} depend on the energy of the system. When the system has a
higher energy, the non-linearity plays a larger role, so energy
essentially plays the role of the coupling constant\cite{AK1}.  
This means that the energy transfer between the different modes
becomes faster, which should lead to quicker pairing\footnote{We thank
  the referee for insight on this point.}.
The time scale for the anharmonic oscillator becomes smaller at higher
energies, which may also lead to shorter pairing times. To study the energy
dependence of the pairing time, we prepared twenty random initial
conditions at various values of $E$, since the pairing times depend on
the initial conditions. The behavior of pairing times, $\tPair$, is
shown in \tabno{pair}, and it can clearly be seen that the pairing
times tend to become shorter for larger values of $E/N$.  It can also
be seen that there is a significant dependence on the initial
condition. In\figno{pairE}, typical pairing behavior is shown for
energies in \tabno{pair}.  Here, only the last exponent pair to sum to
zero is shown for each value of $E/N$. The local Lyapunov exponents
are ordered $\lambda_{1,2,\cdots,N}$ from the exponent with the
largest to the smallest averaged value. In almost all the cases we
studied, the pairing time is smallest for the local Lyapunov exponents
that correspond to the Lyapunov exponents with the largest absolute
values, and vice versa. In the exceptional cases when this ordering is
not followed, the time differences of the ``wrongly'' ordered pairing
times are small.

\begin{figure}[htbp]
  \centering
  \includegraphics[width=8.6cm,clip=true]{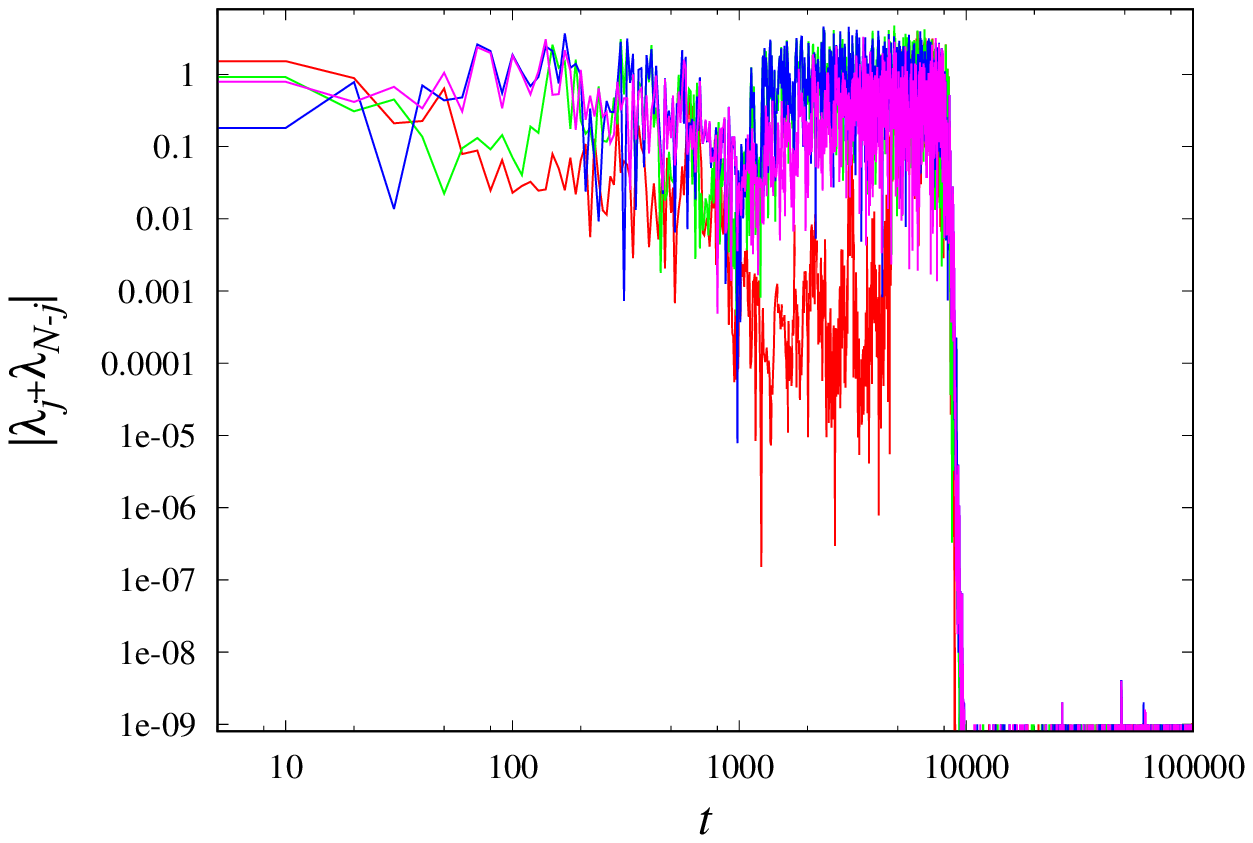}
  \includegraphics[width=8.6cm,clip=true]{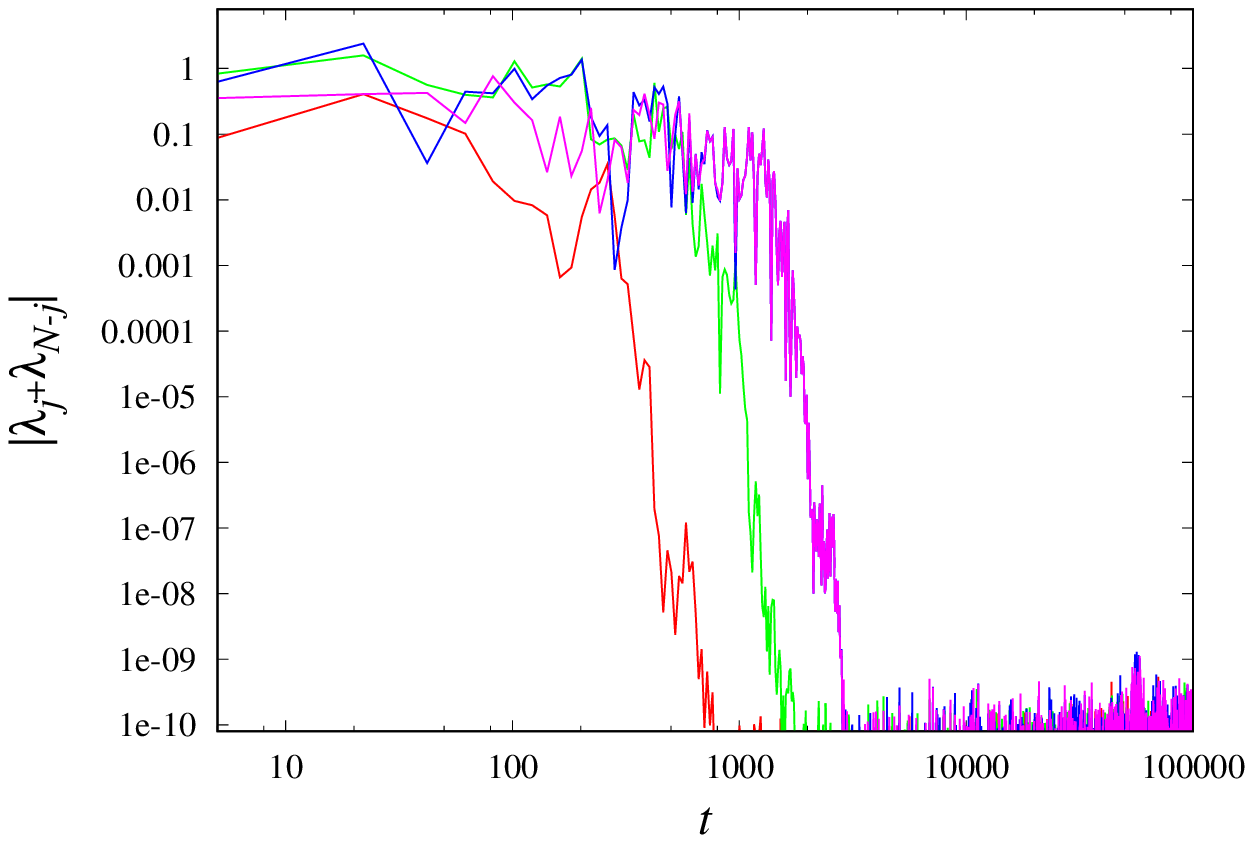}
  \caption{(Left) Pairing behaviors, $|\lambda_j+\lambda_{N-j}|$,
    $j=1$ (red), 2 (green), 3 (blue), 4 (magenta) for $N=4\ \phi^4$
    theory. Initial condition is $q_j=0, (p_j)=(2,2,2,-2)$. (Right)
    The pairing behaviors for a random initial configuration at the
    same energy. We see that the pairing time is much longer for the
    former initial condition.}
  \label{fig:pairE2}
\end{figure}
Let us now investigate the initial conditions for the $\phi^4$ theory,
$N=4$ with periodic boundary conditions, at $E/N=2$.  In
\figno{pairE2}, 
the pairing behavior of the exponents are shown for
the initial condition used in \cite{HooverSnook},
$q_j=0, (p_j)=(2,2,2,-2)$, and for a typical random initial condition
at the same value of $E/N$, for comparison. We see that the specific
initial condition, 
\figno{pairE2}~(left), leads to a much longer
pairing time, $\tPair\sim10^4$, as compared to
$\tPair\sim$few$\times10^3$ for a random initial condition, which is
typical, as can be seen from \tabno{pair}. Another apparent feature is
that the pairing times for all the pairs are essentially the same in
this case, whereas in 
\figno{pairE2}
~(right), they are distinctly
ordered in the decreasing order of the absolute value of the Lyapunov
exponents.

\begin{figure}[htbp]
  \centering
  \includegraphics[width=8.6cm,clip=true]{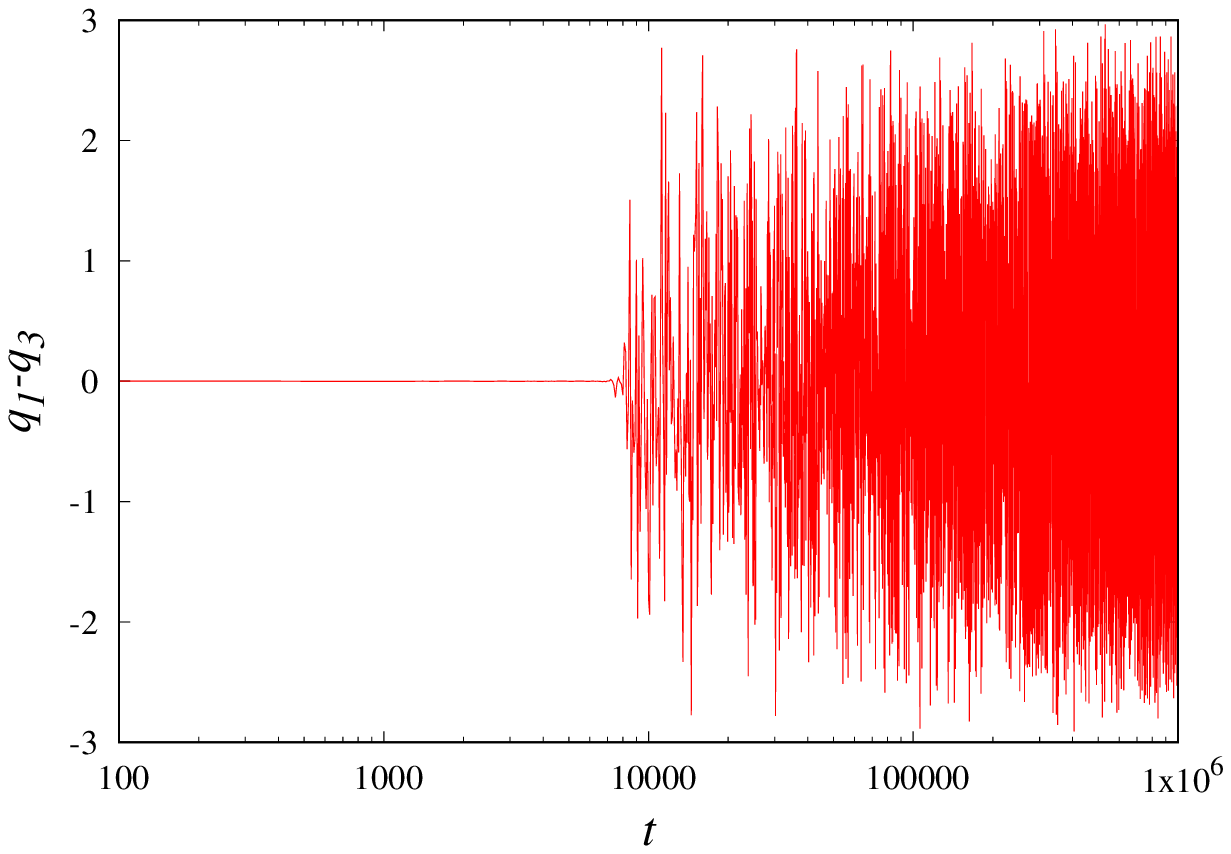}
  \includegraphics[width=8.6cm,clip=true]{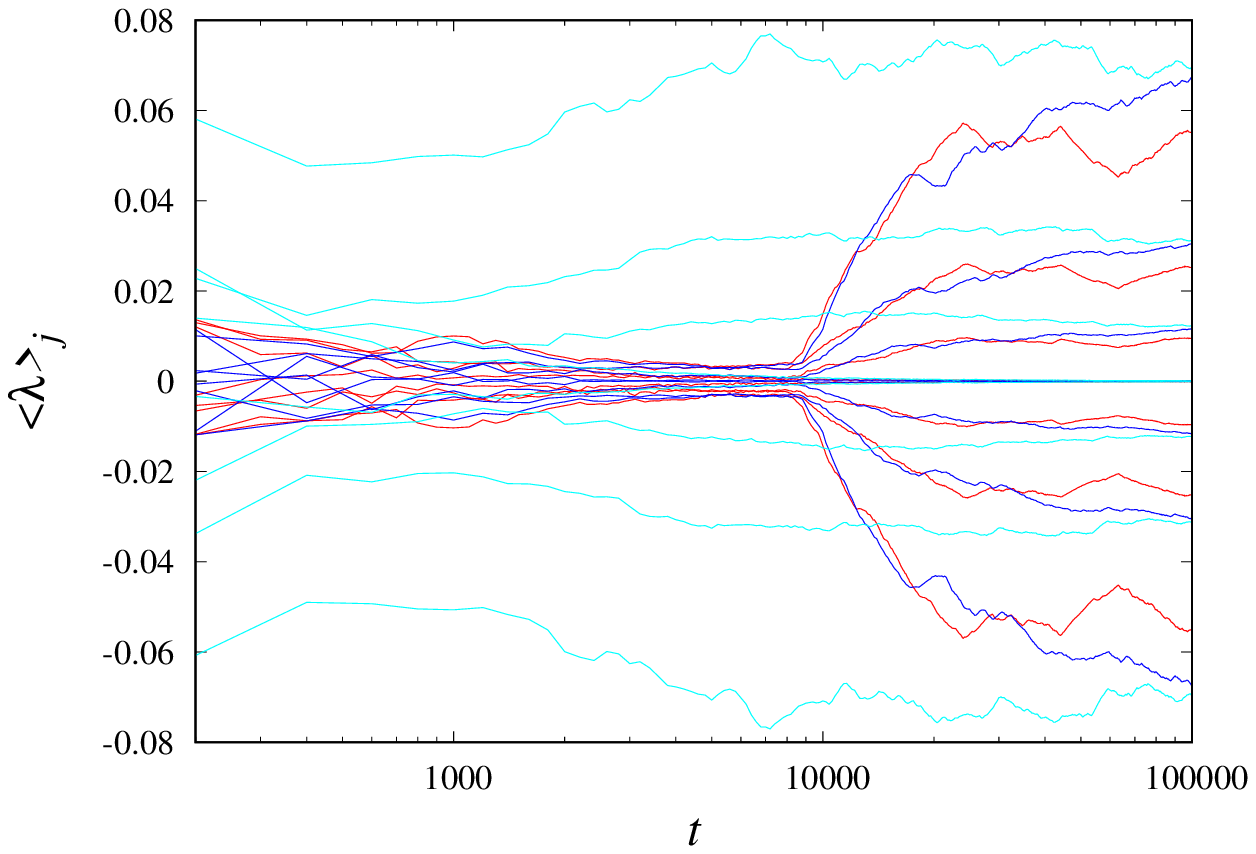}
  \caption{(Left) Dependence of $q_1-q_3$ on time for the initial
    condition $q_j=0, (p_j)=(2,2,2,-2)$ (see text).  (Right) Averaged
    Lyapunov exponents with respect to time. The exponents are shown
    for the initial condition $q_j=0, (p_j)=(2,2,2,-2)$ and the
    tangent vector matrix being the identity matrix, with the
    coordinates ordered as $(q_1,p_1,q_2,p_2,\cdots)$ (red),
    $(q_1,q_2,\cdots,p_1,p_2,\cdots)$ (blue). The spectra for a
    randomly generated initial condition is also shown (cyan), which
    converges more rapidly. }
  \label{fig:pairQ}
\end{figure}
The reason for the above behavior can be understood as follows.  The
initial condition $q_j=0, (p_j)=(2,2,2,-2)$ leads to a chaotic
trajectory, that is restricted to the subspace $q_1=q_3$, as already
discussed in \sect{symmetry} (\figno{trajZ2}). When the trajectory is
restricted to this subspace, pairing does {\it not} occur.  For the fourth
order Runge-Kutta integrator with $dt=10^{-4}$, with the algorithm
computing the Lyapunov spectrum, $q_1=q_3$ starts to break down
visibly at $t\sim7.5\times10^3$, as seen in \figno{pairQ}
~(left). Then the trajectory becomes unrestricted, and within the
additional time of few times $10^3$, the pairing occurs, which is the
typical time for pairing seen in \figno{pairE2} ~(right).  The pairing
times in \figno{pairE2} ~(left) seem similar for all the pairs, since
the difference is roughly an order of magnitude smaller than the time
$q_1=q_3$ relation breaks down. Once the trajectory is not restricted
to the subspace, the pairing occurs, and the Lyapunov exponents
converge, as seen in \figno{pairQ} ~(right).
Computations of the maximal Lyapunov exponent for the trajectory
respecting the symmetry results in $\lambda_1<0.03$ for
$10^4<t\lesssim10^5$\cite{BH3}, which has also been confirmed by our
independent calculations. This would conflict with the results in
\cite{HooverSnook}, and \figno{pairQ}(right) if the symmetry is
unbroken, further adding evidence to the strong possibility that the
symmetry was broken in the previous computation\cite{HooverSnook}.

The above consideration brings up another interesting issue: The
Lyapunov exponents for the initial conditions
$q_j=0, (p_j)=(2,2,2,-2)$, should be those for the trajectory in the
subspace $q_1=q_3$. This is the Lyapunov spectrum that is being
measured up to $t\sim7000$ in 
\figno{pairQ}
~(right). The Lyapunov exponents averaged over subspaces have been
computed for periodic orbits in $\phi^4$ theory, for
example\cite{P4}. However, those spectra exhibit pairing behavior.
This is intuitively satisfying, since the periodic orbits are
symmetric under time reversal. It is unclear if the exponents in the
subspace in this case is unpaired, or not converged enough.  For the
random initial condition in 
\figno{pairE2}
~(right), the
Lyapunov exponents start converging from earlier times, as seen in
\figno{pairQ}
~(right).

\begin{figure}[htbp]
  \centering
  \includegraphics[width=8.6cm,clip=true]{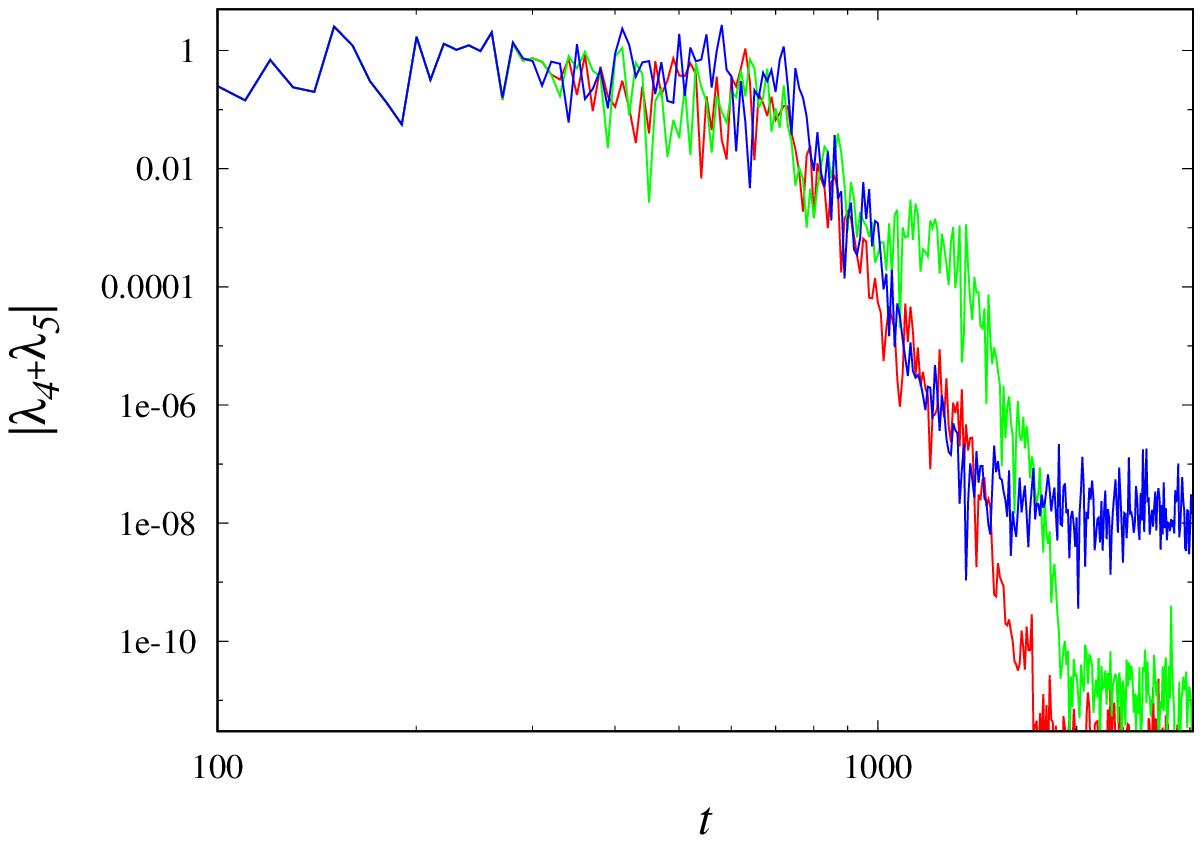}
 \includegraphics[width=8.6cm,clip=true]{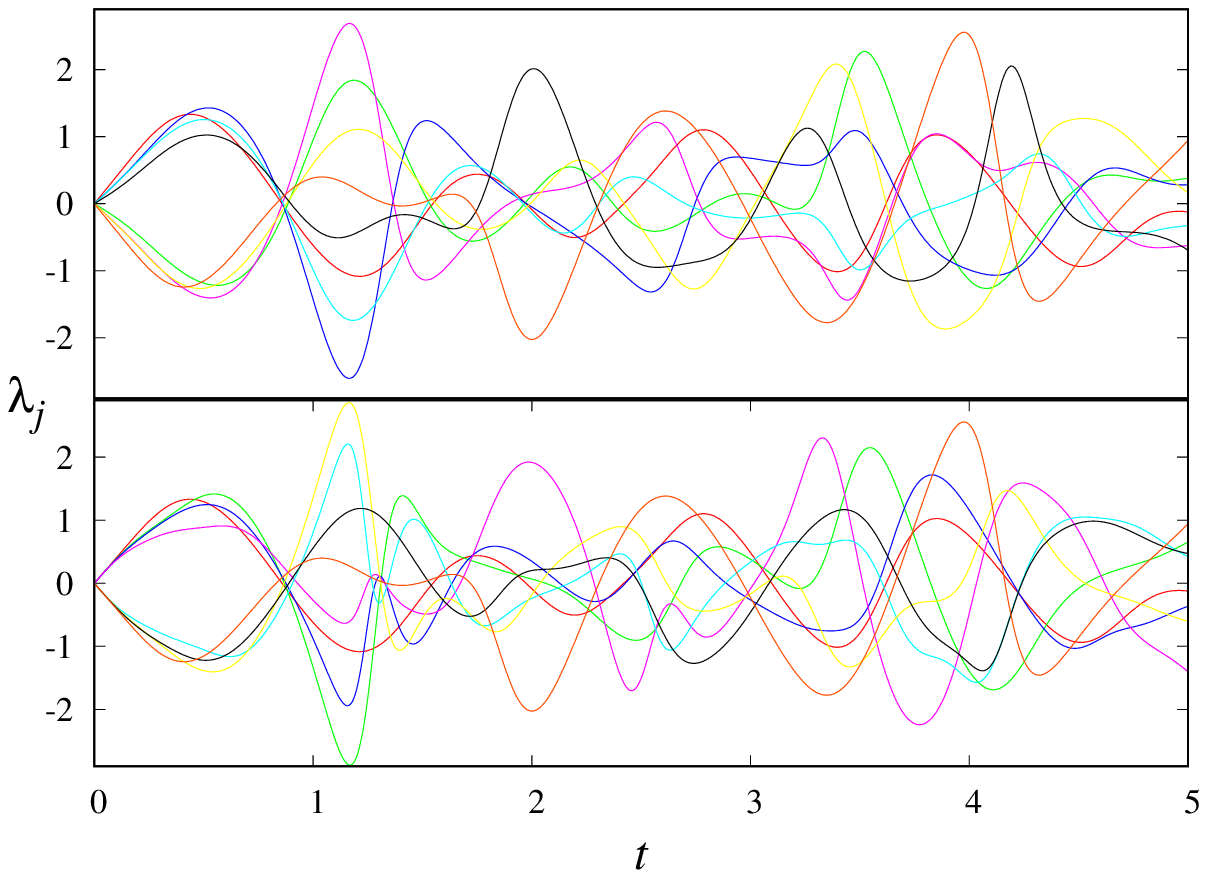}
 \caption{(Left) The pairing, $|\lambda_4+\lambda_5|$, for
   $dt=10^{-5}$ (red), $dt=10^{-4}$ (green), $dt=10^{-3}$ (blue). While
   the pairing is not as precise for larger $dt$, the pairing time
   does not change significantly.  (Right) Local Lyapunov exponents
   for $t=0$ to $t=20$, for the initial condition
   $q_j=0, (p_j)=(2,2,2,-2)$. Top and bottom figures have the
   coordinates ordered as $(q_1,p_1,q_2,p_2,\cdots)$
   $(q_1,q_2,\cdots,p_1,p_2,\cdots)$ in the computation, respectively.
 }
  \label{fig:pairDt}
\end{figure}
Considerations of some technical aspects are in order: The pairing
time can depend on the integrator, or its precision, but in an
indirect way. To obtain Lyapunov exponents, a trajectory (or
trajectories) are integrated over a relatively long time. The
numerical accuracy of tracking the trajectory has limitations,
especially in a dynamical system with chaos. Since the {\em local} Lyapunov
exponent depends on the trajectory within the phase space, its pairing
behavior is also affected. By varying the time step size in the
integrator, $dt$, we can control its precision. We studied how the
pairing times depends on $dt$ in 
\figno{pairDt}. 
Another technical issue is the ordering of the coordinates within the
computation. This affects the local Lyapunov exponents, but does {\it
  not} substantially change the (averaged) Lyapunov exponents, or the
pairing time.  In 
\figno{pairDt}
~(right), local Lyapunov exponents are
shown for the same initial conditions, but with different ordering of
the coordinates. The local Lyapunov exponents are similar, but differ
in their behavior. The averaged behaviors are compared in
\figno{pairQ} 
and are shown to be quite similar. So, the technical
aspects, as far as we have studied, do not influence the pairing
phenomena in an essential way. 
Considering the results above, when computing Lyapunov exponents, it
seems practical to start from a few initial conditions, some perhaps
random, since the pairing times, and consequently the convergence time
for the Lyapunov exponents can vary significantly.

For the properties of Lyapunov exponents, such as pairing, covariant
Lyapunov vectors\cite{CLV1,CLV2,CLVRev} can potentially provide
powerful mathematical tools for their analysis. Their relation to the
orthogonal Gram-Schmidt vectors used in this work are known. While
outside the scope of this paper, it would be interesting to use these
relations to help understand how the pairing times behave. It should
be noted, however, that the covariant Lyapunov vectors themselves
require computation to obtain, so that it is unclear that it provides
an advantage from a practical standpoint.

In this work, we studied symmetry properties, thermodynamic concepts,
and the behavior of local Lyapunov exponents in the $\phi^4$
theory. While seemingly unrelated on the surface, these topics are, as
we have seen, closely intertwined. While clarifying some of the
issues, questions remain, which we feel attests to the depth and the
breadth of the subject.
\acknowledgments We would like to thank William Hoover for his
encouragement, and discussions.  K.A. was supported in part by the
Grant--in--Aid for Scientific Research (\#15K05217) from the Japan
Society for the Promotion of Science (JSPS), and a grant from Keio
University.

\begin{thebibliography}{99}
\bibitem{chaos0}R.C. Hilborn, ``Chaos and nonlinear dynamics'',
  Oxford University Press (New York, 1994).
  \bibitem{chaos1}M. Tabor, ``Chaos and Integrability in Nonlinear
    Dynamics'', John Wiley \& Sons (New York, 1989).
  \bibitem{HH1}W.G. Hoover, C.G. Hoover, ``Simulation and Control of
    Chaotic Nonequilibrium Systems'', World Scientific  Publishing
    Company (Singapore, 2015), and references therein.
\bibitem{HooverSnook}Wm. G. Hoover, C. G. Hoover,
  CMST 23, 73 (2017).
\bibitem{AK1}  K. Aoki and D. Kusnezov, 
  Phys. Lett. A 265, 250 (2000); 
  K. Aoki, D. Kusnezov,  
  Ann. Phys. { 295}, 50 (2002).
\bibitem{Hu}  B. Hu, B. Li, and H. Zhao, 
  Phys. Rev. E { 61}, 3828 (2000).
\bibitem{para}
  Y. Ohnuki and S. Kamefuchi, Phys. Rev. 170, 1279 (1968).
\bibitem{anyon} F. Wilczek, Phys. Rev. Lett. 49, 957 (1982).
\bibitem{orbifold}L.Dixon, J.A.Harvey, C.Vafa, E.Witten,  Nuc. Phys. B  261,  678 (1985).
\bibitem{thooft}  G. {}'t Hooft, Nucl. Phys. B 153, 141  (1979).
  \bibitem{HA1}W.G. Hoover and  K. Aoki, 
  CMST 49,    192 (2017).
  \bibitem{Rosenberg}R. Rosenberg,  
    Adv. Appl. Mech., 9,  156 (1966). 
  \bibitem{NLNM0}
    S.W. Shaw, C. Pierre, 
    Journal of Sound and Vibration  150, 170 (1991), {\it ibid.}  
    164 85 (1993).
  \bibitem{SanduskyPage}K.W. Sandusky and J.B. Page, Phys. Rev. B 50,
    866 (1994).
  \bibitem{Flach}S. Flach, Physica D 91, 223 (1996).
  \bibitem{NLNMRev}    A.F. Vakakis, 
    Mech. Sys.  Sig. Proc. 11, 3 (1997);
    Y.V. Mikhlin and K.V. Avramov
    Appl. Mech. Rev 63, 060802 (2011).
  \bibitem{ChechinRev} T. Bountis, G. Chechin and V. Sakhnenko,
    Int. J. Bif. Chaos 21, 1539 (2011).
 \bibitem{P4} K. Aoki, 
   Phys. Rev. {     E94}, 042209 (2016).
\bibitem{thermo1} S.~Nos\'{e}, J.~Chem.~Phys. {81}, 511 (1984); {
    Mol.~Phys.} {52}, 255 (1984).
\bibitem{thermo2} W.~G.~Hoover, { Phys.~Rev. }{A 31},1695 (1985).
  \bibitem{thermo}F. Reif, ''Fundamentals of Statistical and Thermal
    Physics'', Waveland Press (Long Grove, 2009).
 \bibitem{AK2} K. Aoki, D.
   Kusnezov, 
   Phys. Lett. { A309}, 377 (2003).
  \bibitem{NR}W.H. Press,,‎ S.A. Teukolsky,‎ W.T. Vetterling,
    B.P. Flannery, ``Numerical Recipes: The Art of Scientific
    Computing,  3rd Edition'', Cambridge University Press (New York, 2007).
  \bibitem{Boost} Boost libraries, \url{http://www.boost.org/}.
  \bibitem{Lyap0}
    I. Shimada, T. Nagashima, Prog. Theor. Phys. 61, 1605 (1979).
  \bibitem{Lyap1} 
    G. Benettin, L. Galgani, A. Giorgilli and  J. Strelcyn, Meccanica
    15, 9 (1980); ibid.    15, 21 (1980).
\bibitem{Lyap2} H.A. Posch, W.G.~Hoover, { Phys. Rev. }{ A38}, 473
  (1988). 

\bibitem{BH3}W.G. Hoover, private communication.    
\bibitem{CLV1}V. Oseledets, Tr. Mosk. Mat. Obs. 19, 179 (1968).
\bibitem{CLV2} D. Ruelle,Publ. Math. IHES 50, 275 (1979).
\bibitem{CLVRev}F. Ginelli, H. Chat\'e, R. Livi, A. Politi,
J. Phys. A: Math. Theor. 46, 254005  (2013). 
\end{thebibliography}
\end{document}